\documentclass[showpacs,aps,prd,preprint,nofootinbib,showkeys,unsortedaddress,raggedbottom]{revtex4-1}
\pdfoutput=1

\usepackage{bm}
\usepackage{amsmath}
\usepackage{graphicx}
\usepackage{subfigure}
\usepackage[usenames,dvipsnames]{color}
\definecolor{darkblue}{RGB}{0,0,196}
\usepackage[colorlinks=true,linkcolor=darkblue,citecolor=darkblue,urlcolor=darkblue]{hyperref}

\DeclareMathOperator{\arcsinh}{arcsinh}

\def\be{\begin{equation}}
\def\ee{\end{equation}}

\def\ba{\begin{eqnarray}}
\def\ea{\end{eqnarray}}

\newcommand{\checked}[1]{}

\begin{document}

\title{Anisotropic hydrodynamics for conformal Gubser flow}

\author{Mohammad Nopoush}
\affiliation{Kent State University, Kent OH 44242 USA}
\author{Radoslaw Ryblewski}
\affiliation {The H. Niewodnicza\'nski Institute of Nuclear Physics, Polish Academy of Sciences, PL-31342 Krak\'ow, Poland}
\author{Michael Strickland}
\affiliation{Kent State University, Kent OH 44242 USA}

\begin{abstract}
We derive the equations of motion for a system undergoing boost-invariant longitudinal and azimuthally-symmetric transverse ``Gubser flow'' using leading-order anisotropic hydrodynamics.  This is accomplished by assuming that the one-particle distribution function is ellipsoidally-symmetric in the momenta conjugate to the de Sitter coordinates used to parameterize the Gubser flow.  We then demonstrate that the $SO(3)_q$ symmetry in de Sitter space further constrains the anisotropy tensor to be of spheroidal form.  The resulting system of two coupled ordinary differential equations for the de Sitter-space momentum scale and anisotropy parameter are solved numerically and compared to a recently obtained exact solution of the relaxation-time-approximation Boltzmann equation subject to the same flow.  We show that anisotropic hydrodynamics describes the spatio-temporal evolution of the system better than all currently known dissipative hydrodynamics approaches.  In addition, we prove that anisotropic hydrodynamics gives the exact solution of the relaxation-time approximation Boltzmann equation in the ideal, $\eta/s \rightarrow 0$, and free-streaming, $\eta/s \rightarrow \infty$, limits.
\end{abstract}

\maketitle

\section{Introduction}

Dissipative hydrodynamics is now commonly used to describe the spatio-temporal evolution of the matter created in ultrarelativistic heavy-ion collisions.  The approaches used for this purpose have included relativistic ideal hydrodynamics \cite{Huovinen:2001cy,Hirano:2002ds,Kolb:2003dz}, second-order viscous hydrodynamics \cite{Muronga:2001zk,Muronga:2003ta,Muronga:2004sf,Heinz:2005bw,Baier:2006um,Romatschke:2007mq,Baier:2007ix,Dusling:2007gi,Luzum:2008cw,Song:2008hj,Heinz:2009xj,El:2009vj,PeraltaRamos:2009kg,PeraltaRamos:2010je,Denicol:2010tr,Denicol:2010xn,Schenke:2010rr,Schenke:2011tv,Bozek:2011wa,Niemi:2011ix,Niemi:2012ry,Bozek:2012qs,Denicol:2012cn,Denicol:2012es,PeraltaRamos:2012xk,Calzetta:2014hra,Denicol:2014vaa}, and, most recently, anisotropic hydrodynamics \cite{Martinez:2010sc,Florkowski:2010cf,Ryblewski:2010bs,Martinez:2010sd,Ryblewski:2011aq,Florkowski:2011jg,Martinez:2012tu,Ryblewski:2012rr,Florkowski:2012as,Florkowski:2013uqa,Ryblewski:2013jsa,Bazow:2013ifa,Tinti:2013vba,Florkowski:2014bba,Florkowski:2014txa,Nopoush:2014pfa,Denicol:2014mca}.  For phenomenological applications, second-order viscous hydrodynamics is the most often used dynamical framework, however, traditional viscous hydrodynamics approaches rely on linearization around an isotropic equilibrium state.  If the system has large non-equilibrium corrections, it is not clear a priori that a perturbative treatment will be phenomenologically reliable at all points in spacetime.  In order to address this issue, the framework of anisotropic hydrodynamics was created in order to extend the range of applicability of dissipative hydrodynamics \cite{Martinez:2010sc,Florkowski:2010cf}.  In the anisotropic hydrodynamics framework, the most important (diagonal) components of the energy-momentum tensor are treated non-perturbatively and non-spheroidal/off-diagonal components are treated perturbatively.  This approach has been shown to more accurately describe the evolution of systems subject to boost-invariant and transversely homogeneous (0+1d) flow than traditional viscous hydrodynamics approaches \cite{Florkowski:2013lza,Florkowski:2013lya,Bazow:2013ifa,Florkowski:2014bba,Nopoush:2014pfa}.  

Despite this limited success, to date there is no compelling evidence that anisotropic hydrodynamics can better describe systems that are also expanding transversely.  In this paper, we derive the dynamical equations of anisotropic hydrodynamics for a system subject to Gubser flow \cite{Gubser:2010ze,Gubser:2010ui} and compare them to recently obtained analytic solutions to the Boltzmann equation in the relaxation-time approximation subject to the same flow \cite{Denicol:2014xca,Denicol:2014tha}.  Since Gubser flow includes both cylindrically-symmetric transverse and boost-invariant longitudinal (1+1d) expansion, this will allow us to test the efficacy of anisotropic hydrodynamics in a more realistic setting.  We will also compare to recently obtained solutions using the Israel-Stewart second-order viscous hydrodynamics framework \cite{Marrochio:2013wla} and a complete second-order Grad 14-moment approximation \cite{Denicol:2014tha}.  

In order to implement the anisotropic hydrodynamics framework, we begin by assuming that, to leading order, the one-particle distribution function is ellipsoidally-symmetric in the momenta conjugate to the de Sitter coordinates used to parameterize the Gubser flow and that the argument of the distribution function only depends quadratically on the de Sitter-space momenta.  We then demonstrate that the $SO(3)_q$ symmetry in de Sitter space further constrains the anisotropy tensor to be of spheroidal form.  The resulting system of two coupled ordinary differential equations for the de Sitter-space momentum scale $\hat\lambda$ and anisotropy parameter $\hat\alpha_\varsigma$ are solved numerically and compared to a recently obtained exact solution of the relaxation-time approximation Boltzmann equation subject to Gubser flow.  We show that anisotropic hydrodynamics describes the spatio-temporal evolution of the system better than all currently known dissipative hydrodynamics approaches.  In addition, we prove that anisotropic hydrodynamics gives the exact solution of the relaxation-time approximation Boltzmann equation in the ideal, $\eta/s \rightarrow 0$, and free-streaming, $\eta/s \rightarrow \infty$, limits.

The structure of this paper is as follows.  In Sec.~\ref{sec:conventions} we introduce the conventions used.  In Sec.~\ref{sec:setup} we introduce the Gubser flow profile, de Sitter coordinates, and the Weyl transformation which can be used to make the Gubser flow static.  In Sec.~\ref{sec:bulkvars} we introduce an ellipsoidal ansatz for the de Sitter-space one-particle distribution function and calculate the non-vanishing components of the energy-momentum tensor in de Sitter space based on this.  We then show that requiring $SO(3)_q$ symmetry in de Sitter space constrains the distribution function to be of spheroidal form.  Using this result, in Sec.~\ref{sec:moments} we derive the equations of motion for the de Sitter-space momentum scale and anisotropy parameter by taking the first and second moments of the Boltzmann equation.  In Sec.~\ref{sec:limitingcases} we present the ideal and free-streaming limits of the anisotropic hydrodynamics equations of motion.  In Sec.~\ref{sec:exactsolution} we generalize the exact solution of the relaxation-time-approximation Boltzmann equation subject to Gubser flow obtained originally in Refs.~\cite{Denicol:2014xca,Denicol:2014tha}, to allow for arbitrary anisotropy initial conditions and demonstrate that the anisotropic hydrodynamics equations of motion give the exact result in both the ideal and free-streaming limits.  In Sec.~\ref{sec:results} we compare the numerical solution of the anisotropic hydrodynamics equations of motion with the exact solution and two different viscous hydrodynamics approximations.  Finally, in Sec.~\ref{sec:conclusions} we present our conclusions and an outlook for the future.  We collect some details and ancillary information in four appendices.

\section{Conventions}
\label{sec:conventions}

In this paper, the metric is taken to be ``mostly plus'' such that in Minkowski space with $x^{\mu }=(t,x,y,z)$, the line element is
\be
ds^2 = g_{\mu\nu} dx^\mu dx^\nu = -dt^2 + dx^2 + dy^2 + dz^2 \,.
\ee
Milne coordinates are defined by $\check{x}^\mu=(\tau,x,y,\varsigma)$, where $\tau =\sqrt{t^2 - z^2}$ is the longitudinal proper time, $\varsigma ={\rm tanh}^{-1}(z/t)$ is the longitudinal spacetime rapidity, and the metric is \mbox{$d\check{s}^2=-d\tau^2 + dx^2 + dy^2 + \tau^2d\varsigma^2$}. Since we deal with a system that is cylindrically-symmetric in the lab frame, it is convenient to transform to polar coordinates in the transverse plane with $r=\sqrt{x^{2}{+}y^{2}}$ and $\phi ={\rm tan}^{-1}(y/x)$.  If we use polar coordinates in the transverse plane, we will refer to this as ``polar'' Milne coordinates with components $\tilde{x}^\mu = (\tau,r,\phi,\varsigma)$ and metric $d\tilde{s}^2=-d\tau^2+dr^2 + r^2d\phi^2 + \tau^2d\varsigma^2$.  In some places we denote the scalar product between two 4-vectors with a dot, i.e. $a_\mu b^\mu \equiv a \cdot b$.  In all cases, the flow velocity $u^\mu$ is normalized as $u_{\mu }u^{\mu }=-1$.

\section{Setup}
\label{sec:setup}

Herein we assume that the system is boost invariant and cylindrically symmetric with respect to the beamline at all times.  With this assumption, one can construct a flow with $SO(3)_q{\otimes}SO(1,1){\otimes}Z_2$ symmetry (``Gubser symmetry'') \cite{Gubser:2010ui,Gubser:2010ze}.  In this case, one can show that all dynamical variables depend on $\tau$ and $r$ through the dimensionless combination
\ba 
G(\tau,r)=\frac{1-q^2{\tau}^2+q^2r^2}{2 q{\tau}}\, ,
\label{eq:G}
\ea
where $q$ is an arbitrary energy scale.\footnote{The final results presented herein are expressed in de Sitter space and hold for arbitrary $q$.}  In order to study the dynamics, we start by specifying a basis appropriate for treating a boost-invariant and cylindrically-symmetric system and then simplify the equations of motion by introducing de Sitter coordinates.

\subsection{Boost-invariant and cylindrically-symmetric flow in Minkowski space}

A general tensor basis can be constructed by introducing four 4-vectors in the local rest frame (LRF)
\ba 
u^\mu_{LRF} &\equiv& (1,0,0,0) \, , \notag \\
\mathcal{X}^\mu_{LRF} &\equiv& (0,1,0,0)\, , \notag \\
\mathcal{Y}^\mu_{LRF} &\equiv& (0,0,1,0)\, , \notag \\
\mathcal{Z}^\mu_{LRF} &\equiv& (0,0,0,1)\, .
\ea
The metric in flat spacetime is $g^{\mu\nu}\!=\!\rm diag(-1,\,+1,\,+1,\,+1)$, which can be written in terms of the tensor basis above as
\ba
g^{\mu\nu}=-u^\mu u^\nu+\mathcal{X}^\mu \mathcal{X}^\nu+\mathcal{Y}^\mu \mathcal{Y}^\nu+\mathcal{Z}^\mu \mathcal{Z}^\nu \, .
\label{minkowski-metric}
\ea
For boost-invariant flow which is cylindrically-symmetric around the beamline axis, one can parameterize the basis vectors in the lab frame as
\ba  
u^\mu &=& (\cosh\theta_\perp\cosh\varsigma,\sinh\theta_\perp\cos\phi,\sinh\theta_\perp\sin\phi,\cosh\theta_\perp\sinh\varsigma) , \notag \\
\mathcal{X}^\mu &=& (\sinh\theta_\perp\cosh\varsigma,\cosh\theta_\perp\cos\phi,\cosh\theta_\perp\sin\phi,\sinh\theta_\perp\sinh\varsigma) , \notag \\
\mathcal{Y}^\mu &=& (0,-\sin\phi,\cos\phi,0) ,\notag \\
\mathcal{Z}^\mu &=& (\sinh\varsigma,0,0,\cosh\varsigma)\, . 
\label{eq:minkowski-4vector}
\ea

\subsection{Weyl transformation}
\label{sec:weyl}

In order for a system to be conformally invariant, the dynamics should be invariant under Weyl rescaling \cite{Gubser:2010ui}. A $(m,n)$ tensor of the form $Q^{\mu_1 ...\mu_m}_{\nu_1 ...\nu_n}(x)$ with canonical dimension $\Delta$ transforms under Weyl rescaling as
\ba
 Q^{\mu_1 ...\mu_m}_{\nu_1 ...\nu_n}(x)\,\rightarrow\,\Omega^{\Delta+m-n}Q^{\mu_1 ...\mu_m}_{\nu_1 ...\nu_n}(x)\, ,
\label{eq:weyl-rescaling}
\ea
where $\Omega(x) = \exp(\omega(x))$ with $\omega(x)$ being a function of space and time.
For example, the metric tensor $g_{\mu\nu}$ is a dimensionless tensor of rank 2. Using the relation above with $m=0$, $n=2$, and $\Delta=0$, one finds $[g_{\mu\nu}]=-2$. This means that  $g_{\mu\nu}$ has a conformal weight of $-2$ and transforms under Weyl rescaling as \cite{Gubser:2010ui}
\ba 
g_{\mu\nu} \rightarrow \Omega^{-2}\,g_{\mu\nu}\, .
\label{eq:g-weyl}
\ea

\subsection{Gubser flow and de Sitter coordinates}

The Gubser flow is completely determined by symmetry constraints to be \cite{Gubser:2010ui,Gubser:2010ze}
\ba 
\tilde{u}^\tau &=&\cosh\left[\tanh^{-1} \left(\frac{2q^2\tau r}{1+q^2\tau^2+q^2r^2}\right)\right]\, ,\notag \\
\tilde{u}^r &=&\sinh\left[\tanh^{-1} \left(\frac{2q^2 \tau r}{1+q^2\tau^2+q^2r^2}\right)\right]\, , \notag \\
\tilde{u}^\phi &=& 0\, , \notag \\
\tilde{u}^\varsigma &=& 0 \, .
\label{eq:gubser-flow}
\ea
Using Eqs.~(\ref{eq:minkowski-4vector}) and (\ref{eq:gubser-flow}), one can determine the corresponding transverse rapidity $\theta_\perp$
\be 
\theta_\perp=\tanh^{-1} \left(\frac{2q^2\tau r}{1+q^2\tau^2+q^2r^2}\right).
\label{eq:thetaperp}
\ee
In what follows, we will perform Weyl rescaling and a change of variables to de Sitter coordinates.  For this
purpose, we begin by introducing the de Sitter ``time'' $\rho$ and polar angle $\theta$ \cite{Gubser:2010ui}
\ba 
\sinh{\rho} &=&  - \frac{1-q^2{\tau^2}+q^2r^2}{2q{\tau}}\, ,
\label{eq:desitter1} \\
\tan{\theta} &=& \frac{2qr}{1+q^2{\tau}^2-q^2r^2}\, ,
\label{eq:desitter2}
\ea
where $\tau$ and $r$ are polar Milne 4-vector coordinates $\tilde{x}^\mu=(\tau,\,r,\,\phi,\,\varsigma)$ and $\rho$ and $\theta$ are two of the de Sitter coordinates $\hat{x}^\mu=(\rho,\,\theta,\,\phi,\,\varsigma)$.  Note that, for fixed $r$, the limit $\tau \rightarrow 0^+$ corresponds to the limit $\rho \rightarrow -\infty$ and the limit $\tau\rightarrow\infty$ corresponds to the limit $\rho \rightarrow \infty$.  This means that the de Sitter map covers the future (forward) light cone.

In order to map the flow (\ref{eq:gubser-flow}) to a static one, we follow the prescription of Gubser \cite{Gubser:2010ui,Gubser:2010ze} and make a coordinate transformation combined with a  Weyl rescaling to pass from ${\bf R}^{3,1}$ to $dS_3 \times {\bf R}$ (de Sitter space).  Quantities defined in de Sitter space will be indicated with a hat throughout the paper.  Using Eq.~(\ref{eq:g-weyl}) and the rules for general coordinate transformations of tensors, one can relate the de Sitter-space metric with the Minkowski space metric via
\ba
\hat{g}_{\mu\nu}=\frac{1}{\tau^2}\frac{\partial x^\alpha}{\partial \hat{x}^\mu}\frac{\partial x^\beta}{\partial\hat{x}^\nu}g_{\alpha\beta}\, .
\ea
The de Sitter-space metric tensor in matrix form is
\be
\hat{g}_{\mu\nu} = {\rm diag}(-1,\, \cosh^2\!\rho,\, \cosh^2\!\rho  \sin^2\!\theta ,\, 1) \, ,
\label{eq:desitter-metric}
\ee
and $\hat{g}^{\mu\nu} = \hat{g}^{-1}_{\mu\nu}$, which can also be expressed in terms of tetrads (see (\ref{eq:trans}) below) as $\hat{g}_{\mu\nu}=\hat{u}_\mu\hat{u}_\nu+\hat{\Theta}_\mu\hat{\Theta}_\nu+\hat{\Phi}_\mu\hat{\Phi}_\nu+\hat{\varsigma}_\mu\hat{\varsigma}_\nu$.  The determinant of $\hat{g}_{\mu\nu}$ is
\ba
\hat{g} \equiv \det{\hat{g}_{\mu\nu}} = -\cosh^4\!\rho \sin^2\!\theta\, .
\label{eq:g}
\ea

In order to proceed, we need to establish relations between the Minkowski-space basis vectors and the de Sitter-space basis vectors.  To do this, we first need to know the conformal weights of the Minkowski-space basis vectors.  Knowing that $[g_{\mu\nu}]=-2$ and using $g_{\mu\nu} {\cal X}^\mu {\cal X}^\nu=1$ (where ${\cal X}^\mu$ generally stands for the spacelike Minkowski basis vectors), one concludes that $[{\cal X}^\mu]=1$ and $[{\cal X}_\mu]=-1$. Also, using $g_{\mu\nu} u^\mu u^\nu=-1$, one obtains $[u^\mu]=1$ and $[u_\mu]=-1$. The tensor transformation to relate 4-vectors in de Sitter coordinates to 4-vectors in Minkowski coordinates can be written as follows
\ba 
\hat{u}^\mu &=& \tau\,\frac{\partial \hat{x}^\mu}{\partial x^\nu}\,u^\nu\, , \notag \\
\hat{\Theta}^\mu &=& \tau\,\frac{\partial \hat{x}^\mu}{\partial x^\nu}\,\mathcal{X}^\nu\, , \notag \\
\hat{\Phi}^\mu &=& \tau\,\frac{\partial \hat{x}^\mu}{\partial x^\nu}\,\mathcal{Y}^\nu\, , \notag \\
\hat{\varsigma}^\mu &=& \tau\,\frac{\partial \hat{x}^\mu}{\partial x^\nu}\,\mathcal{Z}^\nu\, .
\label{eq:trans}
\ea
Starting with the Minkowski-space basis vectors in lab frame Eq.~(\ref{eq:minkowski-4vector}), one can use Eq.~(\ref{eq:trans}) and the de Sitter-space identities detailed in Appendix \ref{app:desitterids} to obtain
\ba 
\hat{u}^\mu &=& (1,\,0,\,0,\,0)\, , \notag \\
\hat{\Theta}^\mu &=& (0,\,(\cosh\rho)^{-1},\,0,\,0)\, , \notag \\
\hat{\Phi}^\mu &=& (0,\,0,\,(\cosh\rho \sin\theta )^{-1},\,0)\, , \notag \\
\hat{\varsigma}^\mu &=& (0,\,0,\,0,\,1)\, .
\label{eq:desitter-4vectors}
\ea
One can check explicitly that the orthonormality conditions for the basis vectors are satisfied, i.e.
\ba
\hat{u}\cdot\hat{u}&\equiv & \hat{u}^\mu\hat{u}_\mu=-1\, , \notag \\
\hat{\Theta}\cdot\hat{\Theta}&\equiv & \hat{\Theta}^\mu\hat{\Theta}_\mu=1\, , \notag \\
\hat{\Phi}\cdot\hat{\Phi}&\equiv & \hat{\Phi}^\mu\hat{\Phi}_\mu=1\, , \notag \\
\hat{\varsigma}\cdot\hat{\varsigma}&\equiv & \hat{\varsigma}^\mu\hat{\varsigma}_\mu=1\, ,
\label{eq:u.u}
\ea 
and all other dot products vanish. In de Sitter coordinates, $\theta$ and $\phi$ are transverse coordinates and $\varsigma$ is the longitudinal one.

\subsection{Ellipsoidal form for the distribution function}

We now introduce our ansatz for the one-particle distribution function.  We will assume that, in de Sitter space, the anisotropy tensor is diagonal.\footnote{This is motivated by the fact that in the  linearized viscous hydrodynamics framework one finds that the shear tensor is diagonal in de Sitter space.}  An ellipsoidal anisotropic distribution function can be constructed by introducing a tensor of the form \cite{Nopoush:2014pfa}
\ba 
\hat{\Xi}^{\mu\nu}=\hat{u}^\mu \hat{u}^\nu+\hat{\xi}^{\mu\nu},
\label{eq:aniso-tensor1}
\ea
where $\hat{u}^\mu$ is the four-velocity and $\hat{\xi}^{\mu\nu}$ is a symmetric traceless anisotropy tensor.  Expanding $\hat{\xi}^{\mu\nu}$ in the de Sitter basis gives
\ba
\hat{\xi}^{\mu\nu} = \hat{\xi}_\theta \hat{\Theta}^\mu \hat{\Theta}^\nu+\hat{\xi}_\phi \hat{\Phi}^\mu \hat{\Phi}^\nu+\hat{\xi}_\varsigma \hat{\varsigma}^\mu \hat{\varsigma}^\nu\, .
\label{eq:aniso-tensor2}
\ea
The basis vectors above obey the identities listed in Eq.~(\ref{eq:u.u}).  We require
\ba
\hat{\xi}^{\mu}_{\ \mu} &=& 0 \, , 
\label{eq:tracelessness}
\\
\hat{u}_\mu \hat{\xi}^{\mu\nu} &=& 0 \, .
\ea
Therefore,
\ba
{\hat{\Xi}^\mu}_{\ \mu} = -1 \, ,
\\
\hat{u}_\mu \hat{\Xi}^{\mu\nu} = -\hat{u}^\nu \, .
\ea
Using the tensor $\hat{\Xi}^{\mu\nu}$, one can construct an anisotropic distribution function following Ref.~\cite{Nopoush:2014pfa}\,\footnote{We assume herein that the chemical potential is zero.}
\ba
f(\hat{x},\hat{p})=f_{\rm iso}\left(\frac{1}{\hat\lambda}\sqrt{\hat{p}_\mu\hat{\Xi}^{\mu\nu} \hat{p}_\nu}\right)\, ,
\label{eq:pdf}
\ea
where $\hat\lambda$ can be identified with the de Sitter-space temperature, $\hat{T}$, only when $\hat\xi^{\mu\nu}=0$.

\subsection{Dynamical variables}

Since $\hat\xi^{\mu\nu}$ is traceless and diagonal, one has
\ba
\hat{\xi}_\theta+\hat{\xi}_\phi+\hat{\xi}_\varsigma=0 \, ,
\label{eq:xi-trace}
\ea
which can be verified using Eqs.~(\ref{eq:desitter-metric}) and (\ref{eq:desitter-4vectors}). In order to satisfy $SO(3)_q$ invariance, the distribution function can only depend on $\hat{p}_\Omega^2\equiv\hat{p}_\theta^2+\hat{p}_\phi^2/\!\sin^2\!\theta$ \cite{Denicol:2014tha}. As a result, one must have $\hat{\xi}_\theta=\hat{\xi}_\phi$. Using this, the condition (\ref{eq:xi-trace}) implies
\ba
\hat{\xi}_\theta=-\frac{\hat{\xi}_\varsigma}{2} \, .
\label{eq:xi-trace2}
\ea
For convenience, one can define new parameters $\hat\alpha_i$ as 
\ba
\hat\alpha_i \equiv (1+\hat{\xi}_i)^{-1/2} \, ,
\label{eq:alpha} 
\ea
where $i \in \{\theta,\,\phi,\,\varsigma\}$.  Using Eqs.~(\ref{eq:desitter-metric}),$\,$(\ref{eq:aniso-tensor1}), and (\ref{eq:alpha}) one can simplify the distribution function to
\ba 
f(\hat{x},\hat{p})=f_{\rm iso}\left(\frac{1}{\hat\lambda}\sqrt{\sum_i\frac{\hat{p}^i\hat{p}_i}{\hat\alpha_i^2}}\right)\, ,
\ea
where $\hat{p}_i$ and $\hat{p}^i$ are related through the metric (\ref{eq:desitter-metric}) as before. Note that a trivial consequence of $\hat{\xi}_\theta=\hat{\xi}_\phi$ is
\be
\hat\alpha_\theta = \hat\alpha_\phi \ .
\label{eq:at=af}
\ee

\section{Bulk variables in de Sitter coordinates}
\label{sec:bulkvars}

In order to extract the energy density and pressures from the energy-momentum tensor, one can expand it in a tensor basis (\ref{eq:desitter-4vectors}) in de Sitter coordinates.  Since the distribution function is of ellipsoidal form, the energy-momentum tensor is diagonal in de Sitter space
\ba 
\hat{T}^{\mu\nu} &=& \hat{\varepsilon}\hat{u}^\mu\hat{u}^{\nu}+
\hat{P}_\theta\hat{\Theta}^\mu\hat{\Theta}^{\nu}
+\hat{P}_\phi\hat{\Phi}^\mu\hat{\Phi}^{\nu}+
\hat{P}_\varsigma\hat{\varsigma}^\mu\hat{\varsigma}^{\nu}\, ,
\label{eq:energy-mom}
\ea
where $\hat{\varepsilon}$, $\hat{P}_\theta$, $\hat{P}_\phi$, and $\hat{P}_\varsigma$ are the energy density and pressures in de Sitter coordinates. In the kinetic theory framework, one can use the integral form of $\hat{T}^{\mu\nu}$ to evaluate these quantities. In general, the $n^{\rm th}$-moment of the distribution function is defined as
\ba
&& \hat{\mathcal{I}}^{\mu_1...\mu_n} \equiv \frac{1}{(2\pi)^3}\int\!\!\frac{d^3\hat{p}}{\sqrt{-\hat{g}\,}\hat{p}^0} \, \hat{p}^{\mu_1}...\,\hat{p}^{\mu_n} f(\hat{x},\hat{p})\, ,
\label{eq:nth moments}
\ea
where $\hat{g}$ is defined in Eq.~(\ref{eq:g}).  Taking $n=2$ in Eq.~(\ref{eq:nth moments}), the integral form of the energy-momentum tensor is obtained
\ba
&& \hat{T}^{\mu\nu} \equiv \frac{1}{(2\pi)^3}\int\frac{d^3\hat{p}}{\sqrt{-\hat{g}\,}\hat{p}^0}\hat{p}^\mu \hat{p}^\nu f(\hat{x},\hat{p})\, .
\label{eq:energy-mom-int}
\ea
Taking projections of $\hat{T}^{\mu\nu}$ with the de Sitter-space basis vectors (\ref{eq:desitter-4vectors}), one finds
\ba 
\hat{\varepsilon} &\equiv & \hat{u}_\mu \hat{T}^{\mu\nu} \hat{u}_\nu = \frac{1}{(2\pi)^3} \int\!\!\frac{d^3\hat{p}}{\sqrt{-\hat{g}\,}\hat{p}^\rho} \, \hat{p}^\rho \hat{p}^\rho f_{\rm iso}\left(\frac{1}{\hat\lambda}\sqrt{\sum_i\frac{\hat{p}_i\hat{p}^i}{\hat\alpha^2_i}}\right)\, ,
\label{eq:e-def} \\
\hat{P_\theta} &\equiv & \hat{\Theta}_\mu \hat{T}^{\mu\nu} \hat{\Theta}_\nu = \frac{1}{(2\pi)^3} \int\!\!\frac{d^3\hat{p}}{\sqrt{-\hat{g}\,}\hat{p}^\rho}\cosh^2\!\rho\, \hat{p}^\theta \hat{p}^\theta f_{\rm iso}\left(\frac{1}{\hat\lambda}\sqrt{\sum_i\frac{\hat{p}_i\hat{p}^i}{\hat\alpha^2_i}}\right)\, ,
\label{eq:ept-def} \\
\hat{P_\phi} &\equiv & \hat{\Phi}_\mu \hat{T}^{\mu\nu} \hat{\Phi}_\nu = \frac{1}{(2\pi)^3} \int\!\!\frac{d^3\hat{p}}{\sqrt{-\hat{g}\,}\hat{p}^\rho} \cosh^2\!\rho\sin^2\!\theta\, \hat{p}^\phi \hat{p}^\phi f_{\rm iso}\left(\frac{1}{\hat\lambda}\sqrt{\sum_i\frac{\hat{p}_i\hat{p}^i}{\hat\alpha^2_i}}\right)\, ,
\label{eq:pf-def} \\
\hat{P_\varsigma} &\equiv & \hat{\varsigma}_\mu \hat{T}^{\mu\nu} \hat{\varsigma}_\nu = \frac{1}{(2\pi)^3} \int\!\!\frac{d^3\hat{p}}{\sqrt{-\hat{g}\,}\hat{p}^\rho} \, \hat{p}^\varsigma \hat{p}^\varsigma f_{\rm iso}\left(\frac{1}{\hat\lambda}\sqrt{\sum_i\frac{\hat{p}_i\hat{p}^i}{\hat\alpha^2_i}}\right)\, ,
\label{eq:pv-def}
\ea
where $d^3\hat{p}\equiv d\hat{p}_\theta d\hat{p}_\phi d\hat{p}_\varsigma$ and we have used $\hat{p}^0 = \hat{p}^\rho$. To obtain these results, the following identities were used
\ba 
\hat{u}_\mu \hat{p}^\mu &=& -\hat{p}^\rho\, , \notag \\
\hat{\Theta}_\mu \hat{p}^\mu &=& \hat{p}^\theta\cosh\rho\, , \notag \\
\hat{\Phi}_\mu \hat{p}^\mu &=& \hat{p}^\phi\cosh\rho\sin\theta\, , \notag \\
\hat{\varsigma}_\mu \hat{p}^\mu &=& \hat{p}^\varsigma \, .
\ea
Computing the integrals, the bulk variables in de Sitter coordinates are
\ba 
\hat{\varepsilon}&=&\frac{6\hat\alpha_\theta\hat\alpha_\phi}{(2\pi)^3}\hat\lambda^4 \int_0^{2\pi}d\phi\,\hat\alpha_\perp^2 H_2(y)\, ,
\label{eq:e} \\
\hat{P}_\theta &=&\frac{6\hat\alpha_\theta^3\hat\alpha_\phi}{(2\pi)^3}\hat\lambda^4 \int_0^{2\pi}d\phi\cos^2\!\phi H_{2T}(y)\, , 
\label{eq:pt}\\
\hat{P}_\phi &=&\frac{6\hat\alpha_\theta\hat\alpha_\phi^3}{(2\pi)^3}\hat\lambda^4 \int_0^{2\pi}d\phi\sin^2\!\phi H_{2T}(y)\, , 
\label{eq:pf}\\
\hat{P}_\varsigma &=&\frac{6\hat\alpha_\theta\hat\alpha_\phi}{(2\pi)^3}\hat\lambda^4 \int_0^{2\pi}d\phi\,\hat\alpha_\perp^2 H_{2L}(y)\, .
\label{eq:pv}
\ea
where $\hat\alpha_\perp\equiv\sqrt{\hat\alpha^2_\theta\cos^2\phi+\hat\alpha^2_\phi\sin^2\phi}$, $y \equiv\hat\alpha_\varsigma/\hat\alpha_\perp$, and the $H$-functions are defined as follows
\ba
H_2(y)&\equiv & y\int_{-1}^1 d(\cos\!\theta)\sqrt{y^2\cos^2\!\theta+\sin^2\!\theta}\notag \\ 
&=& \frac{y}{\sqrt{y^2-1}}\left(\tanh^{-1}\!\frac{\sqrt{y^2-1}}{y}+y\sqrt{y^2-1}\right)\, ,
\label{eq:H2}
\ea

\ba
H_{2T}(y)&\equiv & y\int_{-1}^1 \frac{d(\cos\!\theta)\sin^2\!\theta}{\sqrt{y^2\cos^2\!\theta+\sin^2\!\theta}} \notag \\ &=& \frac{y}{(y^2-1)^{3/2}}\left((2y^2-1)\tanh^{-1}\!\frac{\sqrt{y^2-1}}{y}-y\sqrt{y^2-1}\right)\, , 
\label{eq:H2T}
\ea

\ba
H_{2L}(y)&\equiv & y^3\int_{-1}^1 \frac{d(\cos\!\theta)\cos^2\!\theta}{\sqrt{y^2\cos^2\!\theta+\sin^2\!\theta}} \notag \\ &=& \frac{y^3}{(y^2-1)^{3/2}}\left(y\sqrt{y^2-1}-\tanh^{-1}\!\frac{\sqrt{y^2-1}}{y}\right)\, .
\label{eq:H2L}
\ea
As discussed earlier, requiring $SO(3)_q$ invariance in de Sitter space implies (\ref{eq:at=af}).  Together with Eq.~(\ref{eq:xi-trace}), this condition implies that one can write $\hat\alpha_\theta$ in terms of $\hat\alpha_\varsigma$
\ba 
\hat\alpha_\theta = \sqrt{\frac{2\hat\alpha_\varsigma^2}{3\hat\alpha_\varsigma^2-1}} \, .
\label{eq:at}
\ea
Additionally, $\hat\alpha_\phi = \hat\alpha_\theta$ implies that $\hat\alpha_\perp=\hat\alpha_\theta$.  Therefore, one can simplify Eqs.~(\ref{eq:e})-(\ref{eq:pv}) to
\ba
\hat{\varepsilon}&=&\frac{3\,\hat\alpha_\theta^4\hat\lambda^4}{2\pi^2} H_2(\bar{y})\, ,
\label{eq:e2} \\
\hat{P}_\theta &=&\frac{3\,\hat\alpha_\theta^4\hat\lambda^4}{4\pi^2} H_{2T}(\bar{y})\, , 
\label{eq:pt2} \\
\hat{P}_\phi &=&\hat{P}_\theta\, ,
\label{eq:pf2} \\
\hat{P}_\varsigma &=&\frac{3\,\hat\alpha_\theta^4\hat\lambda^4}{2\pi^2} H_{2L}(\bar{y})\, ,
\label{eq:pv2}
\ea
where $\bar{y} \equiv \hat\alpha_\varsigma/\hat\alpha_\theta$ is
\ba
\bar{y}=\sqrt{\frac{3\hat\alpha_\varsigma^2-1}{2}}\, .
\label{eq:ybar}
\ea 

\section{Moments of Boltzmann equation}
\label{sec:moments}

The Boltzmann equation in the relaxation-time approximation in de Sitter coordinates is	
\ba 
\hat{p} \cdot D f &=& \frac{\hat{p}\cdot \hat{u}}{\hat\tau_{\rm eq}}(f-f_{\rm iso})\, ,
\label{eq:boltzmanneq}
\ea
where $D_\mu$ is the covariant derivative defined in Appendix \ref{app:covderiv}, $f_{\rm iso}$ denotes the isotropic equilibrium distribution function, and $\hat\tau_{\rm eq}$ is the relaxation time. Conformal invariance requires that $\hat\tau_{\rm eq}$ is inversely proportional to the temperature, i.e. $\hat\tau_{\rm eq} \propto 1/\hat{T}$. Since we work in the relaxation-time approximation, the exact relation is $\hat\tau_{\rm eq} = 5\hat{\bar\eta}/\hat{T}$, where $\hat{\bar\eta}=\hat\eta/\hat{s}=\eta/s$ with $\hat\eta$ being the Weyl-rescaled shear viscosity and $\hat{s}$ being the  Weyl-rescaled entropy density.  In order to derive the dynamical equations, we take the first and second moments of the Boltzmann equation in de Sitter coordinates.

\subsection{First Moment}

Taking the first moment of the Boltzmann equation (\ref{eq:boltzmanneq}) gives
\ba 
D_\mu \hat{T}^{\mu\nu} &=& 0\, ,
\label{eq:boltzmann1}
\ea
where $\hat{T}^{\mu\nu}$ is the energy-momentum tensor. To obtain (\ref{eq:boltzmann1}) we require that the first moment of the right-hand side of the Boltzmann equation vanishes, so that the energy and momenta are conserved. This results in the so-called dynamical Landau matching condition, which allows us to express the effective temperature $\hat{T}$ in terms of the microscopic parameters
\ba
\hat{T}=\frac{\hat\alpha_\varsigma}{\bar{y}}\left(\frac{H_2(\bar{y})}{2}\right)^{1/4} \hat\lambda\, .
\label{eq:matching-final}
\ea
Using Eqs.~(\ref{eq:energy-mom}) and (\ref{eq:covariant-derivative6}) in Appendix \ref{app:covderiv}, one can expand Eq.~(\ref{eq:boltzmann1}) to obtain
\ba
&&\Gamma^\nu_{\lambda\mu} \hat{T}^{\lambda\mu} + \hat{T}^{\mu\nu} \frac{\partial_\mu\sqrt{-\hat{g}\,}}{\sqrt{-\hat{g}\,}} \notag \\ 
&&+ \hat{u}^{\nu}(\hat{u}^\mu\partial_\mu)\hat{\varepsilon}+\hat{u}^\nu(\partial_\mu\hat{u}^\mu)\hat{\varepsilon}+\hat{\varepsilon}(\hat{u}^\mu\partial_\mu)\hat{u}^\nu \notag \\
&&+ \hat{\Theta}^{\nu}(\hat{\Theta}^\mu\partial_\mu)\hat{P}_\theta+\hat{\Theta}^\nu(\partial_\mu\hat{\Theta}^\mu)\hat{P}_\theta+\hat{P}_\theta(\hat{\Theta}^\mu\partial_\mu)\hat{\Theta}^\nu  \notag \\ 
&&+ \hat{\Phi}^{\nu}(\hat{\Phi}^\mu\partial_\mu)\hat{P}_\phi+\hat{\Phi}^\nu(\partial_\mu\hat{\Phi}^\mu)\hat{P}_\phi+\hat{P}_\phi(\hat{\Phi}^\mu\partial_\mu)\hat{\Phi}^\nu  \notag \\
&&+ \hat{\varsigma}^{\nu}(\hat{\varsigma}^\mu\partial_\mu)\hat{P}_\varsigma+\hat{\varsigma}^\nu(\partial_\mu\hat{\varsigma}^\mu)\hat{P}_\varsigma+\hat{P}_\varsigma(\hat{\varsigma}^\mu\partial_\mu)\hat{\varsigma}^\nu  = 0\, .
\label{eq:boltzman-expand}
\ea
Making use of the de Sitter 4-vectors (\ref{eq:desitter-4vectors}), one can take different projections of Eq.~(\ref{eq:boltzman-expand}).  Taking $\hat{u}_\nu$, $\hat{\Theta}_\nu$, $\hat{\Phi}_\nu$, and $\hat{\varsigma}_\nu$ projections gives, respectively,
\ba
\partial_\rho\hat{\varepsilon} + \tanh\!\rho \,(2\hat{\varepsilon} + \hat{P}_\theta + \hat{P}_\phi) &=& 0\, ,
\label{eq:1th-mom-e}\\
\partial_\theta\hat{P}_\theta +(\hat{P}_\theta-\hat{P}_\phi)\cot\theta &=& 0\, , 
\label{eq:1th-mom-pt}\\
\partial_\phi\hat{P}_\phi &=& 0 \, ,
\label{eq:1th-mom-pf}\\
\partial_\varsigma\hat{P}_\varsigma &=& 0 \, .
\label{eq:1th-mom-pv}
\ea
Using the $SO(3)_q$ symmetry and Eq.~(\ref{eq:at=af}), one can simplify the equations above to
\ba
\partial_\rho\hat{\varepsilon} + 2\tanh\!\rho \,(\hat{\varepsilon} + \hat{P}_\theta) &=& 0\, ,
\label{eq:1th-mom-e2}\\
\partial_\theta\hat{P}_\theta &=& 0\, , 
\label{eq:1th-mom-pt2}\\
\partial_\phi\hat{P}_\phi &=& 0 \, ,
\label{eq:1th-mom-pf2}\\
\partial_\varsigma\hat{P}_\varsigma &=& 0 \, .
\label{eq:1th-mom-pv2}
\ea 
The set of equations above demonstrates that, subject to $SO(3)_q$ symmetry, all fields and physical quantities are functions of $\rho$ exclusively. In other words, the differential equations describing the system reduce to coupled first-order ordinary differential equations, which can be solved by providing initial conditions in de Sitter space. Having the final expressions for $\hat{\varepsilon}$ and $\hat{P}_\theta$, Eqs.~(\ref{eq:e2}) and (\ref{eq:pt2}), one finds the first moment of Boltzmann equation in de Sitter space
\ba
4\frac{d\log\hat\lambda}{d\rho}+\frac{3 \hat\alpha_\varsigma^2\left(\frac{H_{2
   L}(\bar{y})}{H_2(\bar{y})}+1\right)-4 }{3\hat\alpha_\varsigma^2-1}\, \frac{d\log\hat\alpha_\varsigma}{d\rho}+ \tanh\rho\left(\frac{H_{2T}(\bar{y})}{H_2(\bar{y})}+2\right)=0\, .
   \label{eq:1st-mom-final}
\ea

\subsubsection{Equivalence to second-order viscous hydrodynamics}

As a check that our starting point given by Eqs.~(\ref{eq:1th-mom-e})-(\ref{eq:1th-mom-pv}) is consistent with the results obtained previously in the context of second-order viscous hydrodynamics, we can rewrite them in terms of the shear tensor.  To do this, we begin by expanding the shear viscous tensor in terms of the de Sitter-space basis vectors (\ref{eq:desitter-4vectors})
\ba
\hat{\pi}_{\mu\nu} =\hat{\pi}_\theta^\theta\hat{\Theta}_\mu\hat{\Theta}_\nu
+\hat{\pi}_\phi^\phi\hat{\Phi}_\mu\hat{\Phi}_\nu
+\hat{\pi}_\varsigma^\varsigma\hat{\varsigma}_\mu\hat{\varsigma}_\nu\, ,
\label{eq:pi-exp}
\ea
where the different components obey
\ba
\hat{\pi}_\theta^\theta+\hat{\pi}_\phi^\phi+\hat{\pi}_\varsigma^\varsigma=0\, .
\label{eq:pi-condition}
\ea
To proceed, we can use the definition of the shear viscous stress tensor as the correction to the isotropic equilibrium pressures
\be
\hat{P}_i = \hat{P}_{\rm iso}+\hat{\pi}_i^i\, ,\\
\label{eq:pi-def}
\ee
where $i \in \{\theta,\phi,\varsigma\}$.
Using $\hat{P}_{\rm iso}=\hat{\varepsilon}/3$, one obtains
\ba
\hat{P}_\theta+\hat{P}_\phi=\frac{2}{3}\hat{\varepsilon}-\hat{\pi}_\varsigma^\varsigma\, .
\label{eq:pt+pf}
\ea
Substituting Eq.~(\ref{eq:pt+pf}) into Eq.~(\ref{eq:1th-mom-e}) gives
\ba
\partial_\rho\hat{\varepsilon} + \tanh\!\rho \,\left(\frac{8}{3}\hat{\varepsilon}-\hat{\pi}_\varsigma^\varsigma\right) = 0\, .
\ea
Using the thermodynamic relation $\hat{\varepsilon}+\hat{P}_{\rm iso}=\hat{T}\hat{s}$, where $\hat{s}$ is Weyl-rescaled entropy density, one finds $\hat{T}\hat{s}=4\hat{\varepsilon}/3$. In conformal field theory we have $\hat{\varepsilon}\propto \hat{T}^4$. Defining $\bar{\pi}^\varsigma_\varsigma \equiv \hat{\pi}^\varsigma_\varsigma/(\hat{T}\hat{s})$, one obtains the following equation
\ba
\frac{\partial_\rho\hat{T}}{\hat{T}} + \frac{2}{3}\tanh\!\rho = \frac{1}{3}\,\bar{\pi}^\varsigma_\varsigma\tanh\!\rho\, .
\label{eq:denicol}
\ea
This is precisely the same as the first-moment equation obtained originally in Ref.~\cite{Marrochio:2013wla}.

\subsection{Second Moment}

Computing the second moment of Boltzmann equation (\ref{eq:boltzmanneq}) gives
\ba 
D_\lambda\hat{\mathcal{I}}^{\lambda\mu\nu}=-\frac{1}{\hat\tau_{\rm eq}} \left(\hat{u}_\lambda \hat{\mathcal{I}}^{\lambda\mu\nu}_{\rm iso}-\hat{u}_\lambda \hat{\mathcal{I}}^{\lambda\mu\nu}\right) ,
\label{eq:boltzmann2}
\ea
where $\hat{\mathcal{I}}^{\lambda\mu\nu}$ and $\hat{\mathcal{I}}^{\lambda\mu\nu}_{\rm iso}$ can be obtained by taking $n=3$ in Eq.~(\ref{eq:nth moments})
\ba 
\hat{\mathcal{I}}^{\lambda\mu\nu} &=& \int\frac{d^3\hat{p}}{\sqrt{-\hat{g}\,}\hat{p}^0} \, \hat{p}^\lambda \hat{p}^\mu \hat{p}^\nu f(\hat{x},\hat{p})\, , \\
\hat{\mathcal{I}}^{\lambda\mu\nu}_{\rm iso} &=& \int\frac{d^3\hat{p}}{\sqrt{-\hat{g}\,}\hat{p}^0} \, \hat{p}^\lambda \hat{p}^\mu\hat{p}^\nu f_{\rm iso}(\hat{x},\hat{p})\, .
\label{eq:2th-mom-int}
\ea

From the symmetry of the integrands in the definition of $\hat{\mathcal{I}}^{\lambda\mu\nu}$ above, one concludes that $\hat{\mathcal{I}}^{\lambda\mu\nu}$ only contains terms which have an even number of spatial indices. Using the de Sitter-space basis (\ref{eq:desitter-4vectors}), one can expand $\hat{\mathcal{I}}^{\lambda\mu\nu}$ in covariant form as
\ba 
\hat{\mathcal{I}} &\equiv & \hat{\mathcal{I}}_\rho\Big[\hat{u}\otimes \hat{u} \otimes \hat{u} \Big] \notag \\ &&+ \hat{\mathcal{I}}_\theta\Big[\hat{u}\otimes \hat{\Theta} \otimes \hat{\Theta}\,+\,\hat{\Theta}\otimes \hat{u} \otimes \hat{\Theta}\,+\,\hat{\Theta}\otimes \hat{\Theta} \otimes \hat{u} \Big] \notag  \\ &&+ \hat{\mathcal{I}}_\phi\Big[\hat{u}\otimes \hat{\Phi} \otimes \hat{\Phi}\,+\,\hat{\Phi}\otimes \hat{u} \otimes \hat{\Phi}\,+\,\hat{\Phi}\otimes \hat{\Phi} \otimes \hat{u} \Big] \notag \\ &&+ \hat{\mathcal{I}}_\varsigma\Big[\hat{u}\otimes \hat{\varsigma} \otimes \hat{\varsigma}\,+\,\hat{\varsigma}\otimes \hat{u}\otimes \hat{\varsigma}\,+\,\hat{\varsigma}\otimes \hat{\varsigma} \otimes \hat{u} \Big]\, .
\label{2nd-mom-exp}
\ea
For a massless system, one has $\hat{p}^\mu\hat{p}_\mu=0$, which gives the following useful identity
\ba 
\hat{p}^\rho=-\hat{p}_\rho=\sqrt{\frac{\hat{p}_\theta^2}{\cosh^2\!\rho}+\frac{\hat{p}_\phi^2}{\cosh^2\!\rho\sin^2\!\theta}+\hat{p}_\varsigma^2} \; .
\ea
Using the orthonormality relations listed in Eq.~(\ref{eq:u.u}), one can take different projections of $\hat{\mathcal{I}}^{\lambda\mu\nu}$ to obtain the following expressions
\ba \notag
\hat{\mathcal{I}}_\rho &\equiv & -\hat{u}_\lambda\hat{u}_\mu\hat{u}_\nu\hat{\mathcal{I}}^{\lambda\mu\nu} \notag \\ &=& \int d\hat{\mathcal{P}}\hat{p}^2_\rho f_{\rm iso}\!\left(\frac{1}{\hat\lambda}\sqrt{\frac{\hat{p}^2_\theta}{\hat\alpha^2_\theta\cosh^2\!\rho}+\frac{\hat{p}^2_\phi}{\hat\alpha^2_\phi\cosh^2\!\rho\sin^2\!\theta}+\frac{\hat{p}^2_\varsigma}{\hat\alpha^2_\varsigma}}\right) ,
\ea
\ba\notag
\hat{\mathcal{I}}_\theta &\equiv & -\hat{u}_\lambda\hat{\Theta}_\mu\hat{\Theta}_\nu\hat{\mathcal{I}}^{\lambda\mu\nu} \notag \\ &=& \int d\hat{\mathcal{P}}\frac{\hat{p}_\theta^2}{\cosh^2\!\rho}   f_{\rm iso}\!\left(\frac{1}{\hat\lambda}\sqrt{\frac{\hat{p}^2_\theta}{\hat\alpha^2_\theta\cosh^2\!\rho}+\frac{\hat{p}^2_\phi}{\hat\alpha^2_\phi\cosh^2\!\rho\sin^2\!\theta}+\frac{\hat{p}^2_\varsigma}{\hat\alpha^2_\varsigma}}\right) , 
\ea
\ba \notag
\hat{\mathcal{I}}_\phi &\equiv & -\hat{u}_\lambda\hat{\Phi}_\mu\hat{\Phi}_\nu\hat{\mathcal{I}}^{\lambda\mu\nu} \notag \\ &=& \int d\hat{\mathcal{P}}\frac{\hat{p}_\phi^2}{\cosh^2\!\rho\sin^2\!\theta} f_{\rm iso}\!\left(\frac{1}{\hat\lambda}\sqrt{\frac{\hat{p}^2_\theta}{\hat\alpha^2_\theta\cosh^2\!\rho}+\frac{\hat{p}^2_\phi}{\hat\alpha^2_\phi\cosh^2\!\rho\sin^2\!\theta}+\frac{\hat{p}^2_\varsigma}{\hat\alpha^2_\varsigma}}\right) , 
\ea
\ba
\hat{\mathcal{I}}_\varsigma &\equiv & -\hat{u}_\lambda\hat{\varsigma}_\mu\hat{\varsigma}_\nu\hat{\mathcal{I}}^{\lambda\mu\nu} \notag \\ &=& \int d\hat{\mathcal{P}} \hat{p}_\varsigma^2 f_{\rm iso}\!\left(\frac{1}{\hat\lambda}\sqrt{\frac{\hat{p}^2_\theta}{\hat\alpha^2_\theta\cosh^2\!\rho}+\frac{\hat{p}^2_\phi}{\hat\alpha^2_\phi\cosh^2\!\rho\sin^2\!\theta}+\frac{\hat{p}^2_\varsigma}{\hat\alpha^2_\varsigma}}\right) ,
\ea
where 
\ba
d\hat{\mathcal{P}}\equiv\frac{1}{(2\pi)^3}\frac{ d\hat{p}_\theta d\hat{p}_\phi d\hat{p}_\varsigma}{\cosh^2\!\rho\sin\!\theta} \, . 
\ea
The results of the integrals above can be compactly written as
\ba 
&&\hat{\mathcal{I}}_\rho=\hat\alpha\left[\sum_{i=\theta,\phi,\varsigma}\hat\alpha_i^2\right]\hat{\mathcal{I}}_{\rm iso} \, , \label{eq:I-rho}\\
&&\hat{\mathcal{I}}_i=\hat\alpha\hat\alpha_i^2\hat{\mathcal{I}}_{\rm iso} \, ,\label{eq:I-i}
\ea
where $\hat\alpha \equiv \hat\alpha_\theta\hat\alpha_\phi\hat\alpha_\varsigma$ and $\hat{\mathcal{I}}_{\rm iso}\equiv4\hat\lambda^5/\pi^2$.  The coefficients above clearly obey
\ba
\hat{\mathcal{I}}_\rho=\sum_{i=\theta,\phi,\varsigma}\hat{\mathcal{I}}_i \, ,
\label{eq:I-sum}
\ea
which follows from $\hat{g}_{\mu\nu}\hat{\mathcal{I}}^{\mu\nu\lambda}=\hat{g}_{\mu\lambda}\hat{\mathcal{I}}^{\mu\nu\lambda}=\hat{g}_{\nu\lambda}\hat{\mathcal{I}}^{\mu\nu\lambda}=0$ since $\hat{p}^\mu \hat{p}_\mu=0$.  If the system possesses $SO(3)_q$ symmetry, one has
\ba 
\hat{\mathcal{I}}_\theta=\hat{\mathcal{I}}_\phi\, . \label{eq:It=If}
\ea

\subsubsection{Dynamical equations}

Using Eq.~(\ref{2nd-mom-exp}), one can expand Eq.~(\ref{eq:boltzmann2}) as
\ba 
D_\lambda\hat{\mathcal{I}}^{\lambda\mu\nu}&=&\hat{u}^\mu\hat{u}^\nu(\hat{u}^\lambda D_\lambda)\hat{\mathcal{I}}_\rho+
\hat{u}^\mu\hat{u}^\nu(D_\lambda\hat{u}^\lambda)\hat{\mathcal{I}}_\rho +
\hat{\mathcal{I}}_\rho(\hat{u}^\lambda D_\lambda)\hat{u}^\mu\hat{u}^\nu \notag \\
&&+\hat{\Theta}^\mu\hat{\Theta}^\nu (\hat{u}^\lambda D_\lambda)\hat{\mathcal{I}}_\theta
+\hat{\Theta}^\mu\hat{\Theta}^\nu (D_\lambda\hat{u}^\lambda) \hat{\mathcal{I}}_\theta
+\hat{\mathcal{I}}_\theta(\hat{u}^\lambda D_\lambda)\hat{\Theta}^\mu\hat{\Theta}^\nu \notag \\
&&+\hat{u}^\mu\hat{\Theta}^\nu (\hat{\Theta}^\lambda D_\lambda) \hat{\mathcal{I}}_\theta+
\hat{u}^\mu\hat{\Theta}^\nu (D_\lambda\hat{\Theta}^\lambda)\hat{\mathcal{I}}_\theta +
\hat{\mathcal{I}}_\theta(\hat{\Theta}^\lambda D_\lambda)\hat{u}^\mu\hat{\Theta}^\nu \notag \\
&&+\hat{\Theta}^\mu\hat{u}^\nu (\hat{\Theta}^\lambda D_\lambda)\hat{\mathcal{I}}_\theta
+\hat{\Theta}^\mu\hat{u}^\nu (D_\lambda\hat{\Theta}^\lambda)\hat{\mathcal{I}}_\theta
+\hat{\mathcal{I}}_\theta(\hat{\Theta}^\lambda D_\lambda)\hat{\Theta}^\mu\hat{u}^\nu
\notag \\
&&+(\hat{\Theta}\rightarrow\hat{\Phi})+(\hat{\Theta}\rightarrow\hat{\varsigma}) \notag \\
&=&-\frac{1}{\hat\tau_{\rm eq}}\left[\hat{u}_\lambda\hat{\mathcal{I}}^{\lambda\mu\nu}_{\rm iso}-\hat{u}_\lambda\hat{\mathcal{I}}^{\lambda\mu\nu}\right] .
\label{eq:Dynamical-equations}
\ea
Using the identities in Appendices \ref{app:covderiv} and \ref{app:christoffel}, one can simplify Eq.~(\ref{eq:Dynamical-equations}) to
\ba 
D_\lambda\hat{\mathcal{I}}^{\lambda\mu\nu}&&=\hat{u}^\mu\hat{u}^\nu\left[\partial_\rho\hat{\mathcal{I}}_\rho
+2\tanh\rho\,(\hat{\mathcal{I}}_\rho+\hat{\mathcal{I}}_\theta+\hat{\mathcal{I}}_\phi)\right] \notag \\
&&+\hat{\Theta}^\mu \hat{\Theta}^\nu\left[\partial_\rho\hat{\mathcal{I}}_\theta
+4\tanh\rho\,\hat{\mathcal{I}}_\theta\right] \notag \\
&&+\hat{\Phi}^\mu \hat{\Phi}^\nu\left[\partial_\rho\hat{\mathcal{I}}_\phi
+4\tanh\rho\,\hat{\mathcal{I}}_\phi\right] \notag \\
&&+\hat{\varsigma}^\mu \hat{\varsigma}^\nu\left[\partial_\rho\hat{\mathcal{I}}_\varsigma
+2\tanh\rho\,\hat{\mathcal{I}}_\varsigma\right] \notag \\
&&+\frac{\hat{u}^\mu\hat{\Theta}^\nu+\hat{\Theta}^\mu\hat{u}^\nu}{\cosh\rho}\left[\partial_\theta\hat{\mathcal{I}}_\theta+\cot\theta (\hat{\mathcal{I}}_\theta-\hat{\mathcal{I}}_\phi)\right] \notag \\
&&+\frac{\hat{u}^\mu\hat{\Phi}^\nu+\hat{\Phi}^\mu\hat{u}^\nu}{\cosh\rho\sin\theta}\left[\partial_\phi\hat{\mathcal{I}}_\phi\right] \notag \\
&&+\left(\hat{u}^\mu\hat{\varsigma}^\nu+\hat{\varsigma}^\mu\hat{u}^\nu\right)\left[\partial_\varsigma\hat{\mathcal{I}}_\varsigma\right] \notag \\
&&=-\frac{1}{\hat\tau_{\rm eq}}\left[\hat{u}_\lambda\hat{\mathcal{I}}^{\lambda\mu\nu}_{\rm iso}
-\hat{u}_\lambda\hat{\mathcal{I}}^{\lambda\mu\nu}\right]\, .
\ea

From the expression above, we can obtain various scalar projections.  The diagonal projections are:

\vspace{3mm}

{$\hat{u}\hat{u}$-projection}
\ba
\partial_\rho\hat{\mathcal{I}}_\rho
+2\tanh\rho\,(\hat{\mathcal{I}}_\rho+\hat{\mathcal{I}}_\theta+\hat{\mathcal{I}}_\phi)
= \frac{1}{\hat\tau_{\rm eq}}\left[\hat{\mathcal{I}}_{\rho,\rm iso}-\hat{\mathcal{I}}_\rho\right]\, ,
\label{eq:uu-projection}
\ea

{$\hat{\Theta}\hat{\Theta}$-projection}
\ba
\partial_\rho\hat{\mathcal{I}}_\theta + 4\tanh\rho\,\hat{\mathcal{I}}_\theta 
=\frac{1}{\hat\tau_{\rm eq}}\left[\hat{\mathcal{I}}_{\theta,\rm iso}-\hat{\mathcal{I}}_\theta\right]\, ,
\label{eq:tt-projection}
\ea

{$\hat{\Phi}\hat{\Phi}$-projection}
\ba
\partial_\rho\hat{\mathcal{I}}_\phi +4\tanh\rho\,\hat{\mathcal{I}}_\phi 
=\frac{1}{\hat\tau_{\rm eq}}\left[\hat{\mathcal{I}}_{\phi,\rm iso}-\hat{\mathcal{I}}_\phi\right]\, ,
\label{eq:ff-projection}
\ea

{$\hat{\varsigma}\hat{\varsigma}$-projection}
\ba
\partial_\rho\hat{\mathcal{I}}_\varsigma
+2\tanh\rho\,\hat{\mathcal{I}}_\varsigma
=\frac{1}{\hat\tau_{\rm eq}}\left[\hat{\mathcal{I}}_{\varsigma,\rm iso}-\hat{\mathcal{I}}_\varsigma\right]\, .
\label{eq:vv-projection}
\ea
Using Eq.~(\ref{eq:I-sum}), one can verify that equations above are not independent, i.e. Eqs.~(\ref{eq:tt-projection}), (\ref{eq:ff-projection}), and (\ref{eq:vv-projection}) imply that Eq.~(\ref{eq:uu-projection}) is automatically satisfied.  The non-trivial off-diagonal projections are:

\vspace{3mm}

{$\hat{u}\hat{\Theta}$ (or $\hat{\Theta}\hat{u}$)-projection}
\ba 
\partial_\theta\hat{\mathcal{I}}_\theta+\cot\theta(\hat{\mathcal{I}}_\theta-\hat{\mathcal{I}}_\phi)=0\, ,
\label{eq:ut-projection}
\ea

{$\hat{u}\hat{\Phi}$ (or $\hat{\Phi}\hat{u}$)-projection}
\ba 
\partial_\phi\hat{\mathcal{I}}_\phi=0\, ,
\ea

{$\hat{u}\hat{\varsigma}$ (or $\hat{\varsigma}\hat{u}$)-projection}
\ba 
\partial_\varsigma\hat{\mathcal{I}}_\varsigma=0\, .
\ea

All other projections give equations that are trivially satisfied. Using $SO(3)_q$ symmetry, one can use Eq.~(\ref{eq:It=If}) to find the set of independent second-moment equations 
\ba
\partial_\rho\hat{\mathcal{I}}_\theta + 4\tanh\rho\,\hat{\mathcal{I}}_\theta 
&=&\frac{1}{\hat\tau_{\rm eq}}\left[\hat{\mathcal{I}}_{\theta,\rm iso}-\hat{\mathcal{I}}_\theta\right]\, ,
\label{eq:tt-projection2} \\
\partial_\rho\hat{\mathcal{I}}_\varsigma + 2\tanh\rho\,\hat{\mathcal{I}}_\varsigma 
&=&\frac{1}{\hat\tau_{\rm eq}}\left[\hat{\mathcal{I}}_{\varsigma,\rm iso}-\hat{\mathcal{I}}_\varsigma\right]\, ,
\label{eq:vv-projection2} \\
\partial_\theta\hat{\mathcal{I}}_\theta &=& \partial_\phi\hat{\mathcal{I}}_\phi = \partial_\varsigma\hat{\mathcal{I}}_\varsigma = 0\, , \\
\hat{\mathcal{I}}_\rho &=&2\hat{\mathcal{I}}_\theta+\hat{\mathcal{I}}_\varsigma\, , \\
\hat{\mathcal{I}}_{\rho,\rm iso} &=&2\hat{\mathcal{I}}_{\theta,\rm iso}+\hat{\mathcal{I}}_{\varsigma,\rm iso}\, .
\ea
Using Eqs.~(\ref{eq:I-i}), (\ref{eq:tt-projection2}), and (\ref{eq:vv-projection2}), one finds
\ba
\frac{6\hat\alpha_{\varsigma }}{1-3 \hat\alpha _\varsigma ^2}\frac{d \hat\alpha_\varsigma}{d\rho}-\frac{3 \left(3 \hat\alpha_\varsigma^4-4\hat\alpha_\varsigma^2+1\right)}{4\hat\tau_{\rm eq} \hat\alpha _{\varsigma }^5} \left(\frac{\hat{T}}{\hat\lambda}\right)^5+2\tanh\rho=0\, .
\label{eq:2nd-mom-final}
\ea
For the effective temperature appearing above, one uses Eq.~(\ref{eq:matching-final}), which was obtained by requiring energy conservation.

\subsection{Final anisotropic hydrodynamics equations}

Equations~(\ref{eq:1st-mom-final}), (\ref{eq:2nd-mom-final}), and (\ref{eq:matching-final}) form the complete set of equations required in order to describe the de Sitter-space evolution using anisotropic hydrodynamics.  We list them again here in order to provide easier access in the forthcoming discussion
\ba
4\frac{d\log\hat\lambda}{d\rho}+\frac{3 \hat\alpha_\varsigma^2\left(\frac{H_{2
   L}(\bar{y})}{H_2(\bar{y})}+1\right)-4}{3\hat\alpha_\varsigma^2-1} \, \frac{d\log\hat\alpha_\varsigma}{d\rho}+ \tanh\rho\left(\frac{H_{2T}(\bar{y})}{H_2(\bar{y})}+2\right) &=& 0\, ,
\label{eq:1st-mom-final2} \\
\frac{6\hat\alpha_{\varsigma }}{1-3 \hat\alpha _\varsigma ^2} \frac{d \hat\alpha_\varsigma}{d\rho} -\frac{3 \left(3 \hat\alpha_\varsigma^4-4\hat\alpha_\varsigma^2+1\right)}{4\hat\tau_{\rm eq} \hat\alpha _{\varsigma }^5} \left(\frac{\hat{T}}{\hat\lambda}\right)^5+2\tanh\rho=0 \, ,
\label{eq:2nd-mom-final2}
\ea
where $\bar{y} \equiv \hat\alpha_\varsigma/\hat\alpha_\theta = \sqrt{(3\hat\alpha_\varsigma^2-1)/2}$. The $H$-functions appearing above are defined in Eqs.~(\ref{eq:H2})-(\ref{eq:H2L}).  The set of equations can be closed by using the dynamical Landau matching condition
\ba
\hat{T}=\frac{\hat\alpha_\varsigma}{\bar{y}}\left(\frac{H_2(\bar{y})}{2}\right)^{1/4} \hat\lambda .
\label{eq:matching-final2}
\ea

\section{Limiting cases}
\label{sec:limitingcases}

In this section, we consider two limiting cases of Eqs.~(\ref{eq:1st-mom-final2})-(\ref{eq:matching-final2}).  The cases we consider are the ideal ($\hat\tau_{\rm eq}\rightarrow0$) and free-streaming ($\hat\tau_{\rm eq}\rightarrow\infty$) limits. In these two cases, one can dramatically simplify the equations and solve them analytically as first-order ordinary differential equations.  As we will see below, this will allow us to compare our results with the exact solution of Boltzmann equation in the ideal and free-streaming limits, which one can also obtain analytically.

\subsection{Ideal hydrodynamics limit}

In order to take the ideal limit of Eqs.~(\ref{eq:1st-mom-final2}) and (\ref{eq:2nd-mom-final2}), one has to impose the following conditions which require that the system is perfectly isotropic and remains so for all de Sitter time $\rho$
\ba
\hat\alpha_\varsigma\rightarrow1\,, \notag \\
\partial_\rho \hat\alpha_\varsigma\rightarrow0\,, \notag \\
\hat\tau_{\rm eq}\rightarrow0\, .
\ea
With these assumptions, $\bar{y}\rightarrow1$ and $\hat\lambda(\rho)\rightarrow\hat{T}(\rho)$. Using these relations, one finds that Eq.~(\ref{eq:2nd-mom-final2}) is trivially satisfied.  Eq.~(\ref{eq:1st-mom-final2}) simplifies dramatically and can be solved analytically giving
\ba 
\hat{T}(\rho)= \hat{T}_0 \left(\frac{\cosh\rho_0}{\cosh\rho}\right)^{2/3} \, ,
\ea
where $\hat{T}_0 = \hat{T}(\rho_0)$.  This is precisely the solution obtained originally by Gubser and Yarom \cite{Gubser:2010ui}.\footnote{We have generalized the solution to allow the boundary condition to be specified at an arbitrary $\rho_0$.  The form of the Gubser and Yarom solution is recovered when $\rho_0=0$.}

\subsection{Free-streaming limit}

In order to take the free-streaming (FS) limit, one has to take the limit $\hat\tau_{\rm eq}\rightarrow\infty$ of Eqs.~(\ref{eq:1st-mom-final2}) and (\ref{eq:2nd-mom-final2}).  As it turns out, it is also possible to solve the anisotropic hydrodynamics equations analytically in this case.  In this limit, solving Eq.~(\ref{eq:2nd-mom-final2}) gives
\ba
\hat\alpha_\varsigma^2(\rho)&=&\frac{1}{3}+\left(\hat\alpha_{\varsigma,0}^2-\frac{1}{3}\right)\frac{\cosh^2\rho}{\cosh^2\rho_0}\, ,
\ea
where we have specified the boundary condition at $\rho=\rho_0$ and required that $\hat\alpha_\varsigma(\rho_0)=\hat\alpha_{\varsigma,0}$.  With this result, one can obtain an expression for $\bar{y}_{\rm FS}$ using Eq.~(\ref{eq:ybar}) 
\ba
\bar{y}_{\rm FS}=\sqrt{\frac{3\hat\alpha_{\varsigma,0}^2-1}{2}}\frac{\cosh\rho}{\cosh\rho_0}\,.
\ea
Substituting the previous two results into Eq.~(\ref{eq:1st-mom-final2}) and solving it analytically gives
\ba
\hat\lambda(\rho)=\frac{\hat\lambda_0\hat\alpha_{\varsigma,0}}{\hat\alpha_\varsigma(\rho)}\,,
\ea
where we have required $\hat\lambda(\rho_0)=\hat\lambda_0$.  Finally, one can use Eq.~(\ref{eq:matching-final2}) to find the free-streaming limit for the (effective) temperature 
\ba
\hat{T}(\rho)=\hat\lambda_0\hat\alpha_{\varsigma,0}\mathcal{H}_\varepsilon^{1/4}(\mathcal{C}_{\rho_0,\rho}) \, ,
\ea 
where 
\be 
\mathcal{H}_\varepsilon(x) \equiv \frac{x^2}{2}+\frac{x^4}{2}\frac{{\rm tanh}^{-1}\sqrt{1-x^2}}{\sqrt{1-x^2}} \, ,
\ee
and 
\ba
\mathcal{C}_{\rho_0,\rho} \equiv \frac{1}{\bar{y}_{\rm FS}} = \frac{\hat\alpha_{\theta,0}\cosh\rho_0}{\hat\alpha_{\varsigma,0}\cosh\rho}\, .
\ea

Using the effective temperature obtained above, one can find the free-streaming limit of the energy density 
\ba
\hat{\varepsilon}_{\rm FS}=\frac{3\hat\lambda_0^4\hat\alpha_{\varsigma,0}^4}{\pi^2}\mathcal{H}_\varepsilon(\mathcal{C}_{\rho_0,\rho}) .
\label{eq:eFS}
\ea
In addition, one can use Eq.~(\ref{eq:at}) to find $\hat\alpha_\theta(\rho)$. 

Finally, using Eqs.~(\ref{eq:pv2}) and (\ref{eq:pi-def}) one can determine the $\varsigma\varsigma$-component of the viscous stress tensor in the free-streaming limit
\ba 
(\hat{\pi}^\varsigma_\varsigma)_{\rm FS} 
&=& \frac{\hat\lambda_0^4\hat\alpha^4_{\varsigma,0}}{\pi^2}\mathcal{H}_\pi\!\left(\mathcal{C}_{\rho_0,\rho}^{-1}\right),
\label{eq:piFS}
\ea
where 
\ba 
\mathcal{H}_\pi(x)&\equiv &\frac{x\sqrt{x^2-1}(1+2x^2)+(1-4x^2){\rm coth}^{-1}\!\left(x/\sqrt{x^2-1}\right)}{2x^3(x^2-1)^{3/2}}\, .
\ea
We note that the functions $\mathcal{H}_\varepsilon$ and $\mathcal{H}_\pi$ introduced above are closely related to the \mbox{$H$-functions} previously defined in Eq.~(\ref{eq:H2})-(\ref{eq:H2L}) as
\ba
\mathcal{H}_\varepsilon(x)&=&\frac{1}{2}x^4 H_2\!\left(x^{-1}\right)\, , \\
\mathcal{H}_\pi(x)&=&\frac{3}{2x^4}\left(\!H_{2L}(x)-\frac{H_2(x)}{3}\!\right)\, .
\ea

\section{Exact Solution}
\label{sec:exactsolution}

Recently, Denicol et al. obtained an exact solution to the Boltzmann equation subject to Gubser flow in the relaxation-time approximation~\cite{Denicol:2014xca,Denicol:2014tha}.  In order to assess the efficacy of the anisotropic hydrodynamics equations, one can compare the results obtained herein with this exact solution.  We note that one limitation of the exact solution obtained in Refs.~\cite{Denicol:2014xca,Denicol:2014tha} is that the distribution function was assumed to be isotropic at $\rho_0$.  As we will show below, if one assumes that the initial distribution function is of spheroidal form in de Sitter space, then it is possible to allow for an arbitrary pressure anisotropy at $\rho_0$.  This will allow us to compare anisotropic hydrodynamics with the exact solution subject to a variety of different de Sitter-space initial conditions.

In general, the exact solution can be expressed in the form~\cite{Denicol:2014xca,Denicol:2014tha}
\ba
\hat{\varepsilon}(\rho)&=& D(\rho,\rho_0)\hat{\varepsilon}_{\rm FS}+\frac{3}{\pi^2c}\int_{\rho_0}^\rho\!d\rho'D(\rho,\rho')\mathcal{H}_\varepsilon\!\left(\frac{\cosh\rho'}{\cosh\rho}\right)\hat{T}^5(\rho')\, ,
\label{eq:e-exact} \\
\hat{\pi}^\varsigma_\varsigma(\rho)&=& D(\rho,\rho_0)(\hat{\pi}^\varsigma_\varsigma)_{\rm FS}
+ \frac{1}{\pi^2c}\!\int_{\rho_0}^\rho\!d\rho'D(\rho,\rho')\mathcal{H}_\pi\!\left(\frac{\cosh\rho}{\cosh\rho'}\right)\hat{T}^5(\rho')\, ,
\label{eq:pv-exact}
\ea
where 
\begin{equation}
D(\rho_2,\rho_1)=\exp\!\left(-\int_{\rho_1}^{\rho_2} d\rho''\,
\frac{\hat T(\rho'')}{c} \right) .
\label{defineD}
\end{equation}
Above $\hat{T}(\rho) = (\pi^2\hat\varepsilon(\rho)/3)^{1/4}$ is the effective temperature and $c\equiv5\hat\eta/\hat{s}$.  Using Eqs.~(\ref{eq:e-def})-(\ref{eq:pv-def}), $\hat{\varepsilon}_{\rm FS}$ and $(\hat{\pi}^\varsigma_\varsigma)_{\rm FS}$ can be obtained 
\ba
\hat{\varepsilon}_{\rm FS}&\equiv&\frac{1}{(2\pi)^3}\int\!\frac{d^3\hat{p}}{\sqrt{-\hat{g}\,}\hat{p}^\rho}(\hat{p}^\rho)^2 f_{\rm iso}\!\left(\frac{1}{\hat\lambda_0}\sqrt{\frac{\hat{p}^2_\theta}{\hat\alpha^2_{\theta,0}\cosh^2\!\rho_0}+\frac{\hat{p}^2_\phi}{\hat\alpha^2_{\theta,0}\cosh^2\!\rho_0\sin^2\!\theta}+\frac{\hat{p}^2_\varsigma}{\hat\alpha^2_{\varsigma,0}}}\right) \notag \\
&=& \frac{3\hat\lambda_0^4\hat\alpha_{\varsigma,0}^4}{\pi^2}\mathcal{H}_\varepsilon\!\left(\mathcal{C}_{\rho_0,\rho}\right), 
\label{eq:e-exact-f}
\ea
\ba
(\hat{\pi}^\varsigma_\varsigma)_{\rm FS} &\equiv & \frac{1}{(2\pi)^3}\int\!\frac{d^3\hat{p}}{\sqrt{-\hat{g}\,}\hat{p}^\rho}\left(\hat{p}_\varsigma^2-\frac{(\hat{p}^\rho)^2}{3}\right) f_{\rm iso}\!\left(\frac{1}{\hat\lambda_0}\sqrt{\frac{\hat{p}^2_\theta}{\hat\alpha^2_{\theta,0}\cosh^2\!\rho_0}+\frac{\hat{p}^2_\phi}{\hat\alpha^2_{\theta,0}\cosh^2\!\rho_0\sin^2\!\theta}+\frac{\hat{p}^2_\varsigma}{\hat\alpha^2_{\varsigma,0}}}\right) \notag \\
&=& \frac{\hat\lambda_0^4\hat\alpha^4_{\varsigma,0}}{\pi^2}\mathcal{H}_\pi\left(\mathcal{C}_{\rho_0,\rho}^{-1}\right). 
\label{eq:eq:pv-exact-f}
\ea
By using the results above, the integral equations (\ref{eq:e-exact}) and (\ref{eq:pv-exact}) allow for an arbitrary momentum-space anisotropy at $\rho=\rho_0$ with $\hat\alpha_\varsigma(\rho_0)=\hat\alpha_{\varsigma,0}$ and $1/3<\hat{\alpha}_{\varsigma,0}^2<\infty$. In the original work \cite{Denicol:2014xca,Denicol:2014tha}, the solutions obtained were restricted to the case $\hat\alpha_{\varsigma,0}=1$.  If one takes $\hat\alpha_{\varsigma,0}=1$, the expressions above reduce to the ones obtained in \cite{Denicol:2014xca,Denicol:2014tha}.  Importantly, we find that Eqs.~(\ref{eq:eFS}) and (\ref{eq:piFS}) correspond precisely to the exact free-streaming limits obtained above.  This means that, if the initial distribution function at $\rho_0$ is of spheroidal form in de Sitter space, anisotropic hydrodynamics gives the exact solution in the free-streaming limit.

\section{Numerical Results}
\label{sec:results}

In the general case, it is necessary to solve Eqs.~(\ref{eq:1st-mom-final2}), (\ref{eq:2nd-mom-final2}), and (\ref{eq:matching-final2}) numerically.  Since they are ordinary first-order differential equations, this task is rather straightforward.  In order to complete the solution, however, we need to specify a boundary condition.  
This might be non-trivial task since not all choices lead to physical results.  As shown in Appendix B of Ref.~\cite{Denicol:2014tha}, in the exact solution, some initial conditions can result in complex-valued temperatures, etc.  While such solutions may be mathematically sound, they are clearly not physical.  However, as discussed in Appendix B of Ref.~\cite{Denicol:2014tha}, if one fixes the boundary condition on the ``left'' ($\rho \rightarrow -\infty$), which corresponds to the ``distant past'' in de Sitter time, one has freedom to choose the initial condition.  In addition, with this boundary condition, one can smoothly take the limit $\eta/s \rightarrow 0$ in order to obtain the ideal hydrodynamics result (see Fig. 8 of Ref.~\cite{Denicol:2014tha}).  This limit is not guaranteed for other choices of $\rho_0$.  

More importantly, we want to specify a set of initial conditions on a fixed proper-time surface $\tau=\tau_0$ and then take the limit $\tau_0 \rightarrow 0^+$ so that we can describe the system's evolution in the entire forward light cone.  For this reason, in what follows we will always fix the boundary condition on the left.  As discussed above, these boundary conditions will also allow us to smoothly go from the ideal to free-streaming limits unambiguously.  In practice, specifying numerical boundary conditions at extremely large negative $\rho$ and obtaining the full solution also for positive $\rho$ is time-consuming, particularly for the exact solution that we intend to compare with.  For this reason, we will present solutions in which the boundary condition is fixed at a large, but finite, negative $\rho$.  In all plots shown, we fix the boundary condition at $\rho_0 = -10$.

\begin{figure}[t]
\hspace{-7mm}\includegraphics[width=1.04\linewidth]{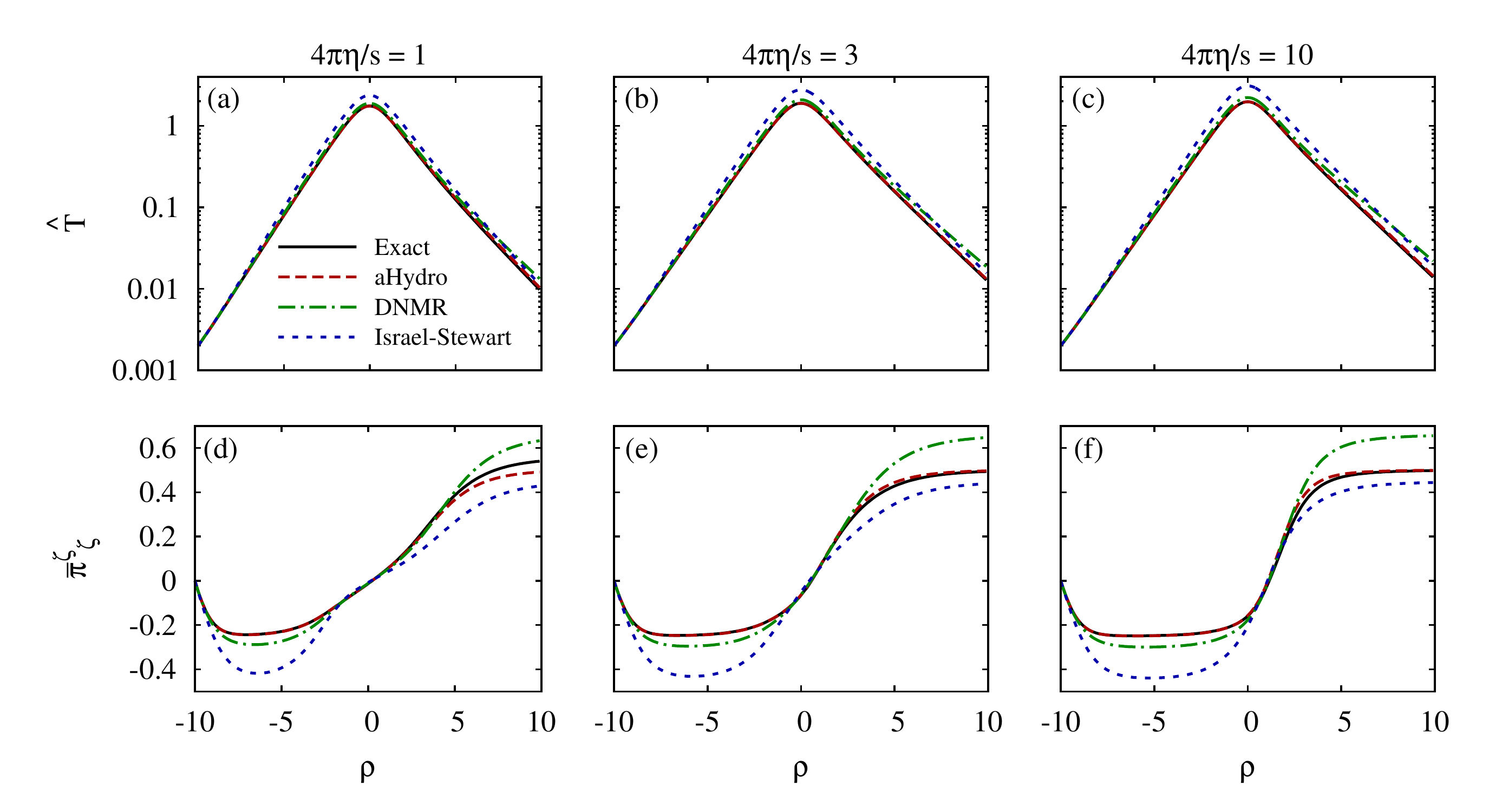}
\vspace{-12mm}
\caption{In the top row, we compare the de Sitter-space effective temperature $\hat{T}$ obtained from the exact solution (black solid line), the anisotropic hydrodynamics equations obtained herein (red dashed line), the DNMR second-order approach (green dot-dashed line), and the Israel-Stewart second-order approximation (blue dotted line).  The columns from left to right correspond to three different choices of the shear viscosity to entropy density ratio with $4 \pi \eta/s \in \{1,3,10\}$, respectively.  In the bottom row, we compare results for the scaled shear $\bar{\pi}_{\varsigma}^{\varsigma} \equiv \hat{\pi}_\varsigma^\varsigma/(\hat{T}\hat{s})$.  The labeling and values of $4 \pi \eta/s$ in the bottom row are the same as in the top row.  In all cases, at $\rho=\rho_0=-10$, we fixed the initial effective temperature to be $\hat{T}_0 = 0.002$ and the initial anisotropy to be  $\hat\alpha_{\varsigma,0}=1$, which corresponds to an isotropic initial condition in de Sitter space.
}
\label{fig:fig-rho-10-T0p002-az1}
\end{figure}

\begin{figure}[t]
\hspace{-7mm}\includegraphics[width=1.04\linewidth]{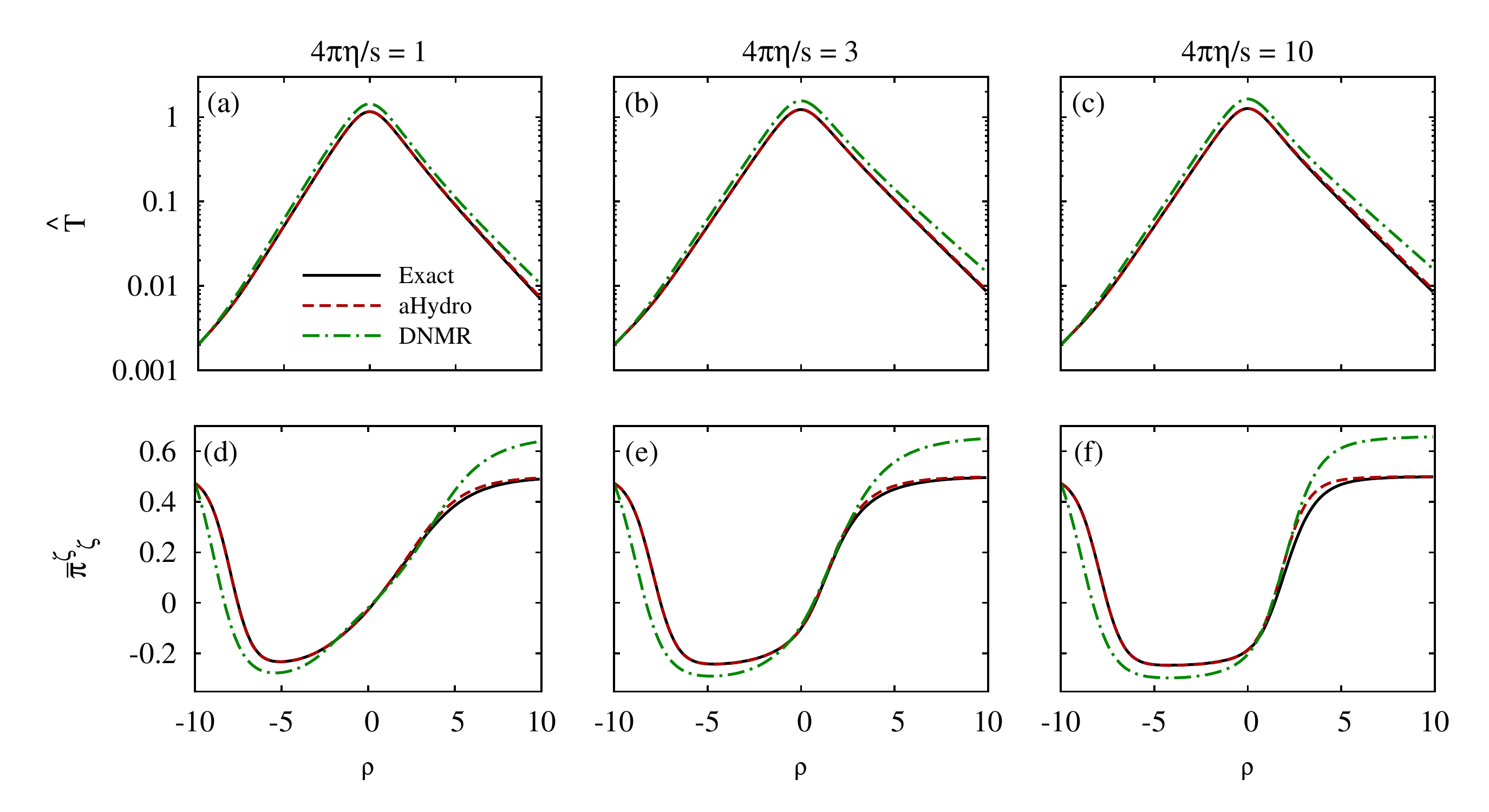}
\vspace{-12mm}
\caption{Same as Fig.~\ref{fig:fig-rho-10-T0p002-az1} except here we take $\hat{\alpha}_{\varsigma,0}=10$, which corresponds to a prolate initial condition in de Sitter space.  We do not include the Israel-Stewart approximation result, because, for this boundary condition, the Israel-Stewart equations are unstable and diverge in the negative-$\rho$ region.}
\label{fig:fig-rho-10-T0p002-az10}
\end{figure}

\begin{figure}[t]
\hspace{-7mm}\includegraphics[width=1.04\linewidth]{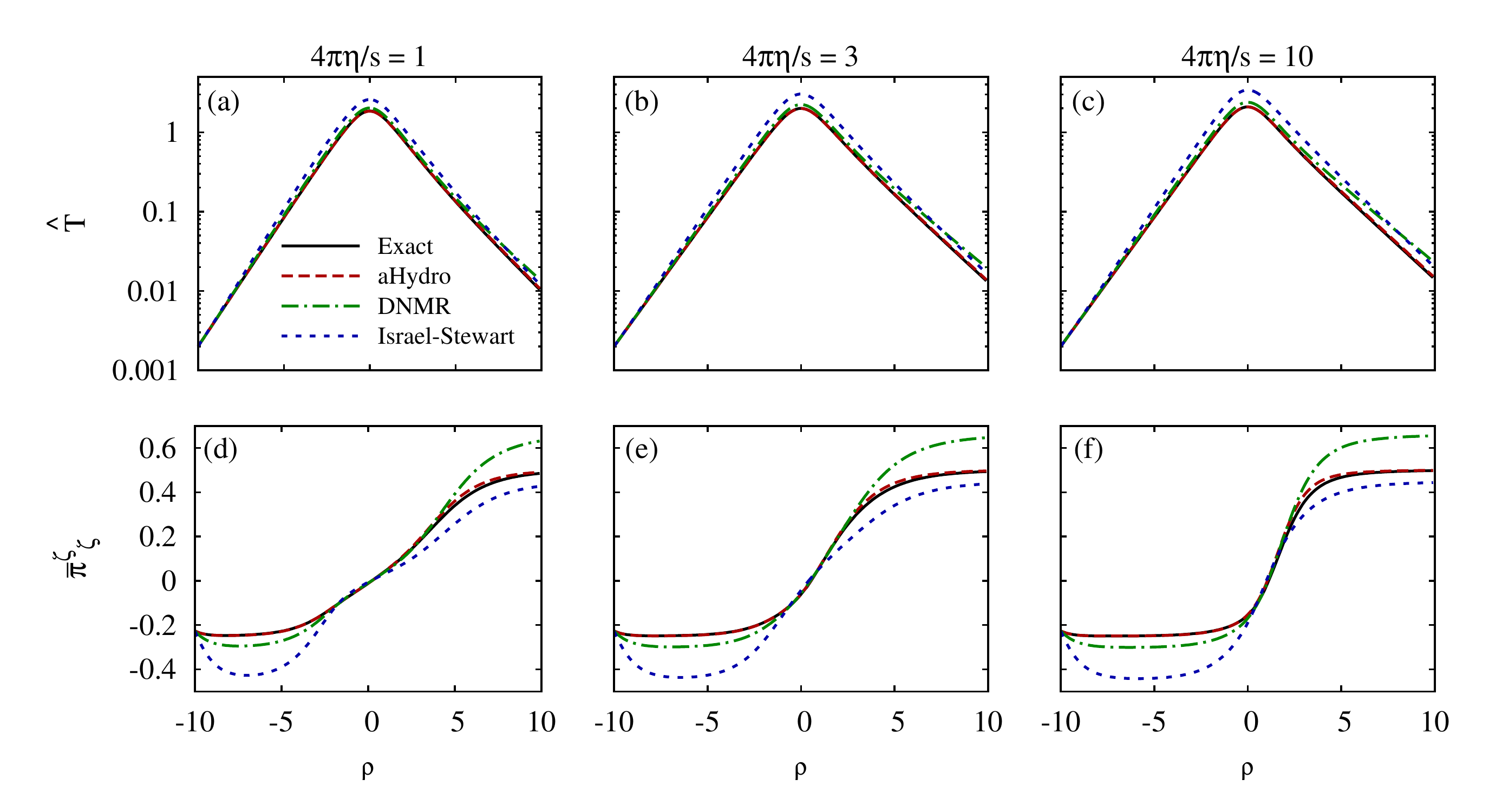}
\vspace{-12mm}
\caption{Same as Fig.~\ref{fig:fig-rho-10-T0p002-az1} except here we take $\hat{\alpha}_{\varsigma,0}=0.6$, which corresponds to an oblate initial condition in de Sitter space.}
\label{fig:fig-rho-10-T0p002-az0p6}
\end{figure}

In addition to comparing to the generalization of the exact result of Refs.~\cite{Denicol:2014xca,Denicol:2014tha}, we will also compare with results obtained using the Israel-Stewart second-order viscous hydrodynamics approximation \cite{Marrochio:2013wla} and a complete second-order Grad 14-moment approximation \cite{Denicol:2014tha}.  For the second-order hydrodynamic approximations, one has to 
solve two coupled ordinary differential equations subject to a boundary condition at $\rho = \rho_0$. For the Israel-Stewart (IS) case, the necessary equations are~\cite{Marrochio:2013wla}
\ba
&&\frac{1}{\hat{T}}\frac{d\hat{T}}{d\rho }+\frac{2}{3}\tanh \rho =
\frac{1}{3}\bar{\pi}_{\varsigma }^{\varsigma }(\rho )\,\tanh \rho \,, 
\label{eq:istemp}
\\
&&\frac{d\bar{\pi}_{\varsigma }^{\varsigma }}{d\rho }
+\frac{4}{3}\left( \bar{\pi}_{\varsigma }^{\varsigma }\right)^{2}\tanh \rho
+\frac{\bar{\pi}_\varsigma^\varsigma}{\hat{\tau}_\pi} =\frac{4}{15}\tanh\rho \,,
\label{eq:isshear}
\ea
where $\bar{\pi}_{\varsigma}^{\varsigma} \equiv \hat{\pi}_\varsigma^\varsigma/(\hat{T}\hat{s})$ and $\hat{\tau}_\pi = 5\hat\eta/(\hat{s}\hat{T})$. As mentioned above, one can go beyond the IS approximation presented in Ref.~\cite{Marrochio:2013wla} and also include the complete second-order contribution (see Appendix A of Ref.~\cite{Denicol:2014tha} for further details). In this case, the second equation above should be replaced by
\begin{equation}
\frac{d\bar{\pi}_{\varsigma }^{\varsigma }}{d\rho }
+\frac{4}{3}\left(\bar{\pi}_{\varsigma }^{\varsigma }\right) ^{2}\tanh \rho 
+\frac{\bar{\pi}_\varsigma^\varsigma}{\hat{\tau}_\pi} 
= \frac{4}{15} \tanh \rho 
+ \frac{10}{21}\bar{\pi}_\varsigma^\varsigma \tanh\rho \,.
\label{eq:dnmrshear}
\end{equation}
If Eq.~(\ref{eq:dnmrshear}) is used, the result is labeled as DNMR.

In Figs.~\ref{fig:fig-rho-10-T0p002-az1}-\ref{fig:fig-rho-10-T0p002-az0p6} we present our numerical solutions of Eqs.~(\ref{eq:1st-mom-final2}), (\ref{eq:2nd-mom-final2}), and (\ref{eq:matching-final2}) and compare the results to the exact solution and the two second-order viscous hydrodynamics approximations.  In these three figures we take $\hat\alpha_{\varsigma,0}=1, 10,$ and $0.6$, which correspond to an initially isotropic, prolate ($\hat{P}_\theta = \hat{P}_\phi < \hat{P}_\varsigma$), and oblate  ($\hat{P}_\theta = \hat{P}_\phi > \hat{P}_\varsigma$) initial condition, respectively.  In all cases, at $\rho=\rho_0=-10$, we fixed the initial effective temperature to be $\hat{T}_0 = 0.002$.  In the top row of all three figures, we compare the de Sitter-space effective temperature $\hat{T}$ obtained from the exact solution (black solid line), the anisotropic hydrodynamics equations obtained herein (red dashed line), the DNMR second-order approach (green dot-dashed line), and the Israel-Stewart second-order approach (blue dotted line).  The columns from left to right correspond to three different choices of the shear viscosity to entropy density ratio with $4 \pi \eta/s \in \{1,3,10\}$, respectively.  In the bottom row of all three figures, we compare results for the scaled shear $\bar{\pi}_{\varsigma}^{\varsigma} \equiv \hat{\pi}_\varsigma^\varsigma/(\hat{T}\hat{s})$.  The labeling and values of $4 \pi \eta/s$ in the bottom row are the same as in the top row.  

As can be seen from Figs.~\ref{fig:fig-rho-10-T0p002-az1}-\ref{fig:fig-rho-10-T0p002-az0p6}, the anisotropic hydrodynamics equations obtained herein provide the best approximation to the exact result in all cases.  For the temperature, it is very difficult to distinguish the anisotropic hydrodynamics result from the exact result.  For the scaled shear $\bar{\pi}_{\varsigma}^{\varsigma} \equiv \hat{\pi}_\varsigma^\varsigma/(\hat{T}\hat{s})$, there are visible differences between the anisotropic hydrodynamics solutions and the exact solution in the region between $\rho \gtrsim  0$ for small $\eta/s$, but at large $\rho$ one sees that anisotropic hydrodynamics has the correct asymptotic behavior.\footnote{Using Eqs.~(\ref{eq:1st-mom-final2})-(\ref{eq:matching-final2}), one finds that in the limit $\rho \rightarrow \infty$, $\hat\alpha_\varsigma \sim \exp(\rho/3)$ and $\hat\lambda \sim \exp(-2\rho/3)$.  As a consequence, one finds that the anisotropic hydrodynamics equations give $\bar{\pi}_{\varsigma}^{\varsigma} = 0.5$ in the limit $\rho \rightarrow \infty$ independent of the value of $\hat\tau_{\rm eq}$.}

Between the two hydrodynamic approximations, we find that, for negative $\rho$, the DNMR solutions better reproduces the exact solution for $\bar{\pi}_{\varsigma}^{\varsigma} \equiv \hat{\pi}_\varsigma^\varsigma/(\hat{T}\hat{s})$, whereas for positive $\rho$ the IS solution seems to perform better overall.  That being said, we find that in the range of de Sitter times considered, the DNMR solution better reproduces the exact solution for the effective temperature.

\section{Conclusions}
\label{sec:conclusions}

In this paper we have used the framework of anisotropic hydrodynamics to derive two coupled ordinary differential equations that describe the evolution of the de Sitter-space scale parameter $\hat\lambda$ and anisotropy parameter $\hat\alpha_{\varsigma}$.  Our final analytic results are listed in Eqs.~(\ref{eq:1st-mom-final2}), (\ref{eq:2nd-mom-final2}), and (\ref{eq:matching-final2}).  Using these equations we could find the evolution of the effective temperature $\hat{T}$ and shear correction $\bar{\pi}_{\varsigma}^{\varsigma}$ in de Sitter time.  We demonstrated that these equations reproduce both the ideal ($\eta/s=0$) and free-streaming ($\eta/s \rightarrow \infty$) limits of the exact solution obtained in Ref.~\cite{Denicol:2014tha}.  In order to make a more general comparison, we extended the exact solution of Ref.~\cite{Denicol:2014tha} to allow for arbitrary momentum-space anisotropy in the de Sitter-space initial condition.  Our numerical results indicate that Eqs.~(\ref{eq:1st-mom-final2}), (\ref{eq:2nd-mom-final2}), and (\ref{eq:matching-final2}) provide an excellent approximation to the exact solution and, hence, this work provides further evidence that the anisotropic hydrodynamics approximation   might provide a superior approximation even when including transverse expansion.  That being said, the transverse flow pattern considered herein (``Gubser'' flow) is rather special, and we cannot generalize beyond the specific case studied herein to a general transverse flow at this point in time.

In the numerical results section we presented solutions for the de Sitter-space evolution of the effective temperature and scaled shear.  The solutions obtained herein can be easily mapped back to Milne space, giving the full spatio-temporal evolution for a boost-invariant and cylindrically-symmetric system for arbitrary values of parameter $q$, which sets the spatial extent of the solution.  Using this mapping, one can obtain the radial temperature profile at any given proper time.  This can be used as an initial condition for subsequent evolution in Milne space.  In a forthcoming paper, we plan to compare the recently obtained 1+1d ellipsoidal anisotropic hydrodynamics equations of Tinti and Florkowski \cite{Tinti:2013vba} with both the anisotropic hydrodynamics equations obtained herein and the exact solution.  Additionally, it will be interesting to see if the methods used herein can also describe the effects of vorticity following Refs.~\cite{Hatta:2014gqa,Hatta:2014gga}.

\acknowledgments{
We thank G.~Denicol, M.~Martinez, U.~Heinz, and J.~Noronha for useful discussions.  R.R. was supported by 
Polish National Science Center Grant No.~DEC-2012/07/D/ST2/02125.  M.S. was supported by U.S. DOE Award No.~\rm{DE-AC0205CH11231}.
}

\appendix

\section{De Sitter coordinates identities}
\label{app:desitterids}

In this appendix, we present some useful identities and derivatives of the de Sitter coordinates which we have used in our calculations. As mentioned in Eqs.~(\ref{eq:desitter1}) and (\ref{eq:desitter2}), de Sitter coordinates are defined as 
\ba 
{\rho}(\tau,r) &=&  \arcsinh\left(-\frac{1-q^2{\tau^2}+q^2r^2}{2q{\tau}}\right) ,
 \\
{\theta}(\tau,r) &=& \arctan\left(\frac{2qr}{1+q^2{\tau}^2-q^2r^2}\right) .
\ea
Taking partial derivatives and using Eq.~(\ref{eq:thetaperp}) for $\theta_\perp$, one can obtain the necessary derivatives of $(\rho,\theta)$ with respect to $(\tau,r)$
\ba
\frac{\partial\rho}{\partial\tau}&=& q \, (\cosh\rho-\sinh\rho\cos\theta)\, , \notag \\
 \frac{\partial\rho}{\partial r}&=&- q \sin\theta \, , \notag \\
 \frac{\partial\theta}{\partial\tau}&=&-q \frac{\sin\theta}{\cosh\rho} \, , \notag \\
 \frac{\partial\theta}{\partial r}&=& q\,(1-\cos\theta\tanh\rho)\, ,
\ea
and inversely
\ba 
\frac{\partial\tau}{\partial\rho}&=&\tau\cosh\theta_\perp\, , \notag \\ 
\frac{\partial\tau}{\partial\theta}&=&q\tau r\, , \notag \\
\frac{\partial r}{\partial\rho}&=&\tau\sinh\theta_\perp\, , \notag \\
\frac{\partial r}{\partial\theta}&=&q\tau r\coth\theta_\perp\, .
\ea
Note that the variables above are also related through the following useful relations
\ba
\tau&=&\frac{1}{q}\frac{\sinh\theta_\perp}{\sin\theta}\, , \\
r&=&\frac{1}{q}\cosh\rho\sinh\theta_\perp\, . 
\ea

\section{The covariant derivative}
\label{app:covderiv}

The covariant derivative is the generalization of the directional derivative of a vector field which acts as a derivative along tangent vectors of a manifold.  Its action on an arbitrary scalar $\varphi$ and rank-1 and rank-2 tensors (indicated by $V$ below) is 
\ba
D_\mu \varphi &=&\partial_\mu \varphi\, , \\
\label{eq:covariant-derivative1} 
D_\mu V_\nu &=&\partial_\mu V_\nu-\Gamma^\lambda_{\beta\mu}V_\lambda\, , \\
\label{eq:covariant-derivative2}
D_\mu V_{\alpha\beta}&=&\partial_\mu V_{\alpha\beta}-\Gamma^\lambda_{\alpha\mu}V_{\lambda\beta}
-\Gamma^\lambda_{\beta\mu}V_{\lambda\alpha}\, ,\\
\label{eq:covariaCovariantnt-derivative3}
D_\mu V^\nu &=&\partial_\mu V^\nu+\Gamma^\nu_{\mu\lambda}V^\lambda\, , \\
\label{eq:covariant-derivative4}
D_\mu V^\mu &=&\frac{1}{\sqrt{-g\,}}\partial_\mu\left(\sqrt{-g\,}V^\mu\right)\, , \\
\label{eq:covariant-derivative5}
D_\mu V^{\mu\nu}&=&\frac{1}{\sqrt{-g\,}}\partial_\mu\left(\sqrt{-g\,}V^{\mu\nu}\right)+\Gamma^\nu_{\lambda\mu} V^{\lambda\mu}\, ,\\
\label{eq:covariant-derivative6}
D_\lambda V^{\mu\nu}&=&\partial_\lambda V^{\mu\nu}+\Gamma^\mu_{\lambda\eta} V^{\eta\nu}+\Gamma^\nu_{\lambda\eta} V^{\mu\eta}\, ,
\label{eq:covariant-derivative7}
\ea
where $\Gamma^\nu_{\mu\lambda}$ are Christoffel symbols, which are
\ba
\Gamma^\nu_{\mu\lambda}=\frac{1}{2}g^{\nu\sigma}(\partial_\mu g_{\sigma\lambda}+\partial_\lambda g_{\sigma\mu}-\partial_\sigma g_{\mu\lambda})\, .
\label{eq:christoffel}
\ea

\section{Christoffel symbols in de Sitter coordinates}
\label{app:christoffel}

Starting from Eq.~(\ref{eq:christoffel}) and using the de Sitter metric (\ref{eq:desitter-metric}),  one obtains the following non-vanishing Christoffel symbols 
\ba
\Gamma^\rho_{\theta\theta}&=&\sinh\rho\cosh\rho\, , \\
\Gamma^\rho_{\phi\phi}&=&\sin^2\theta\sinh\rho\cosh\rho\, , \\
\Gamma^\theta_{\rho\theta}&=&\Gamma^\theta_{\theta\rho}=\tanh\rho\, , \\
\Gamma^\theta_{\phi\phi}&=&-\sin\theta\cos\theta\, , \\
\Gamma^\phi_{\rho\phi}&=&\Gamma^\phi_{\phi\rho}=\tanh\rho\, , \\
\Gamma^\phi_{\theta\phi}&=&\Gamma^\phi_{\phi\theta}=\cot\theta\, .
\ea

\section{The anisotropy tensor in different coordinate systems}
\label{app:xitrans}

In this appendix, we present the transformation of the anisotropy tensor from de Sitter to Milne and polar Milne coordinates.  The tensors in the different cases are indicated by $\hat{\xi}^\mu_\nu$, $\tilde{\xi}^\mu_\nu$, and $\check\xi^\mu_\nu$ in de Sitter, polar Milne, and Milne coordinates, respectively. According to Sec.~\ref{sec:weyl}, since $\hat{\xi}^\mu_\nu$ is a dimensionless tensor of rank 2 with one up and one down index, it has a conformal weight of 0. Therefore, 
\ba
\tilde{\xi}^\mu_\nu=\frac{\partial \tilde{x}^\mu}{\partial\hat{x}^\alpha}\frac{\partial\hat{x}^\beta}{\partial\tilde{x}^\nu}\hat{\xi}^\alpha_\beta\, .
\ea
The anisotropy tensor in de Sitter space is expanded using Eqs.~(\ref{eq:aniso-tensor2}). Using Eq.~(\ref{eq:desitter-4vectors}), one can expand it in matrix form as 
\ba
&&\hat{\xi}^\mu_\nu=
  \begin{pmatrix}
     0 & 0 & 0 & 0 \\ 0 & \hat{\xi}_\theta & 0 & 0\\  0 & 0 & \hat{\xi}_\phi & 0 \\ 0 & 0 & 0 & \hat{\xi}_\varsigma \end{pmatrix} .
\ea
Using the derivative relations in App.~\ref{app:desitterids}, one can find the matrix forms of $\tilde{\xi}^\mu_\nu$ and $\check{\xi}^\mu_\nu$
\ba
&&\tilde{\xi}^\mu_\nu=
  \begin{pmatrix}
     -\hat{\xi}_\theta\sinh^2\theta_\perp & \hat{\xi}_\theta\sinh\theta_\perp\cosh\theta_\perp & 0 & 0 \\-\hat{\xi}_\theta\sinh\theta_\perp\cosh\theta_\perp & \hat{\xi}_\theta\cosh^2\theta_\perp & 0 & 0\\  0 & 0 & \hat{\xi}_\phi
 & 0 \\ 0 & 0 & 0 & \hat{\xi}_\varsigma \end{pmatrix} ,
\ea
\be
\!\check\xi^\mu_\nu\!=\! 
  \begin{pmatrix}
     \!-\hat{\xi}_\theta\sinh^2\theta_\perp & \hat{\xi}_\theta\frac{\sinh(2\theta_\perp\!)}{2}\cos\phi & \hat{\xi}_\theta\frac{\sinh(2\theta_\perp\!)}{2}\sin\phi & 0 \\\!-\hat{\xi}_\theta\frac{\sinh(2\theta_\perp\!)}{2}\cos\phi &\, \hat{\xi}_\theta\cosh^2\!\theta_\perp\cos^2\!\phi+\hat{\xi}_\phi\sin^2\!\phi & (\hat{\xi}_\theta\cosh^2\!\theta_\perp\!- \hat{\xi}_\phi)\sin\!\phi\cos\!\phi & 0\!\\ \!-\hat{\xi}_\theta\frac{\sinh (2\theta_\perp\!)}{2}\sin\!\phi &  (\hat{\xi}_\theta\cosh^2\!\theta_\perp\!-\!\hat{\xi}_\phi)\sin\!\phi\cos\!\phi & \hat{\xi}_\theta\cosh^2\!\theta_\perp\!\sin^2\!\phi\!+\!\hat{\xi}_\phi\cos^2\!\phi\!
 & 0 \\ 0 & 0 & 0 & \hat{\xi}_\varsigma \end{pmatrix} \! ,
\ee
where, in  $\tilde{\xi}^\mu_\nu$ the indices are taken from $(\mu,\nu) \in \{\tau,r,\phi,\varsigma\}$, and in $\check\xi^\mu_\nu$ they are taken from $(\mu,\nu) \in \{\tau,x,y,\varsigma\}$.  Since we started with the basis vectors in Minkowski space lab frame, one needs to boost them to find their form in the local rest frame (LRF).\footnote{In both cases, only a transverse boost is required.  For the case of polar Milne coordinates, one can make a pure radial boost.}  Constructing the necessary boost from the fluid velocity 4-vector appropriate to each coordinate system one finds
\ba
&&\left(\tilde{\xi}^\mu_\nu\right)_{\rm LRF}=
  \begin{pmatrix}
     0 & 0 & 0 & 0 \\ 0 & \hat{\xi}_\theta & 0 & 0 \\ 0 & 0 & \hat{\xi}_\phi & 0 \\ 0 & 0 & 0 & \hat{\xi}_\varsigma \end{pmatrix} ,
\ea
\ba
&&\left(\check\xi^\mu_\nu\right)_{\rm LRF}=
  \begin{pmatrix}
     0 & 0 & 0 & 0 \\ 0 & \frac{\hat{\xi}_++\hat{\xi}_-\cos(2\phi)}{2} & \hat{\xi}_-\sin\phi\cos\phi & 0 \\ 0 & \hat{\xi}_-\sin\phi\cos\phi & \frac{\hat{\xi}_+-\hat{\xi}_-\cos(2\phi)}{2} & 0 \\ 0 & 0 & 0 & \hat{\xi}_\varsigma \end{pmatrix} ,
\ea
where $\hat{\xi}_+\equiv\hat{\xi}_\theta+\hat{\xi}_\phi$ and $\hat{\xi}_-\equiv\hat{\xi}_\theta-\hat{\xi}_\phi$. Using $SO(3)_q$ symmetry, one finds $\hat{\xi}_+\rightarrow 2\hat{\xi}_\theta$ and $\hat{\xi}_-\rightarrow0$ and, therefore,
\ba
&&\left(\tilde{\xi}^\mu_\nu\right)_{\rm LRF}=
  \begin{pmatrix}
     0 & 0 & 0 & 0 \\ 0 & \hat{\xi}_\theta & 0 & 0 \\ 0 & 0 & \hat{\xi}_\theta & 0 \\ 0 & 0 & 0 & \hat{\xi}_\varsigma \end{pmatrix} ,
\ea
\ba
&&\left(\check\xi^\mu_\nu\right)_{\rm LRF}=
  \begin{pmatrix}
     0 & 0 & 0 & 0 \\ 0 & \hat{\xi}_\theta & 0 & 0 \\ 0 & 0 & \hat{\xi}_\theta & 0 \\ 0 & 0 & 0 & \hat{\xi}_\varsigma \end{pmatrix} .
\ea
From the results above, one can conclude that the LRF anisotropy tensor in polar Milne coordinates is diagonal, irrespective of whether the system is $SO(3)_q$-symmetric or not, however, in Milne coordinates, the anisotropy is only diagonal if the system is $SO(3)_q$-symmetric.

\bibliography{desitter}

\end{document}